\documentclass[sigplan,screen]{acmart} 

\usepackage{booktabs}   
\usepackage{subcaption} 

\usepackage{latexsym}
\usepackage{amssymb}
\usepackage{graphicx}
\usepackage{stmaryrd}
\usepackage{tikz}
\usetikzlibrary{automata,positioning}

\usepackage[english]{babel}
\usepackage[T1]{fontenc}
\usepackage{lscape}
\usepackage{listings}
\usepackage{fancyhdr}

\newcommand{\bi}{\begin{array}[t]{@{}l@{}}}
\newcommand{\ei}{\end{array}}
\newcommand{\ba}{\begin{array}}
\newcommand{\ea}{\end{array}}
\newcommand{\bda}{\[\ba}
\newcommand{\eda}{\ea\]}
\newcommand{\bp}{\begin{quote}\tt\begin{tabbing}}
\newcommand{\ep}{\end{tabbing}\end{quote}}

\newcommand{\tlabel}[1]{\mbox{(#1)}}

\newcommand{\turns}{\, \vdash \,}

\newcommand{\sgap}{\quad}
\newcommand{\bgap}{\quad\quad}

\newcommand{\mathem}{\sf}
\newcommand{\IN}{\mbox{\mathem in}}
\newcommand{\LET}{\mbox{\mathem let}}

\newcommand{\CASE}{\mbox{\mathem case}}

\newcommand{\ELSE}{\mbox{\mathem else}}
\newcommand{\IF}{\mbox{\mathem if}}
\newcommand{\OF}{\mbox{\mathem of}}

\newcommand{\CLASS}{\mbox{\mathem class}}

\newcommand{\WHILE}{\mbox{\mathem while}}
\newcommand{\THROW}{\mbox{\mathem throw}}
\newcommand{\TRY}{\mbox{\mathem try}}
\newcommand{\CATCH}{\mbox{\mathem catch}}
\newcommand{\NEW}{\mbox{\mathem new}}
\newcommand{\THIS}{\mbox{\mathem this}}
\newcommand{\NULL}{\mbox{\mathem null}}
\newcommand{\JOIN}{\mbox{\mathem join}}
\newcommand{\EXCEPTION}{\mbox{\mathem exception}}
\newcommand{\OBJ}{\mbox{\mathem obj}}

\newcommand{\tr}[1]{}
\newcommand{\nottr}[1]{}
\newcommand{\fd}[0]{\mathit{fd}}
\newcommand{\cd}[0]{\mathit{cd}}
\newcommand{\md}[0]{\mathit{md}}
\newcommand{\vd}[0]{\mathit{vd}}

\newcommand{\FD}[0]{\mathit{FD}}
\newcommand{\CD}[0]{\mathit{CD}}
\newcommand{\MD}[0]{\mathit{MD}}
\newcommand{\VD}[0]{\mathit{VD}}

\newcommand{\figuretwo}[1]
        {\begin{minipage}{8.5cm} #1 \end{minipage}}

\newcommand{\figtwocol}[3]
        {\begin{figure}
            \figuretwo{#3 \hrule \vspace{-3mm} \caption{\label{#1}#2} } \end{figure}}

\newcommand{\arrow}{\rightarrow}

\newcommand{\ignore}[1]{}
\newcommand{\pjs}[1]{}
\newcommand{\jw}[1]{}

\newcommand{\PHI}{\mbox{\mathem phi}}

\newcommand{\RET}{\mbox{return}}

\newcommand{\loc}[1]{{\tt loc}(#1)}

\newcommand{\ssameth}[1]{{\mathbb M\mathbb D}_{\tt ssa}\llbracket#1\rrbracket}
\newcommand{\ssaexp}[1]{{\mathbb E}_{\tt ssa}\llbracket#1\rrbracket}

\newcommand{\ssavdecl}[1]{{\mathbb V\mathbb D}_{\tt ssa}\llbracket#1\rrbracket}
\newcommand{\ssablk}[1]{{\mathbb B}_{\tt ssa}\llbracket#1\rrbracket}
\newcommand{\ssablks}[1]{{\overline{\mathbb B}}_{\tt ssa}\llbracket#1\rrbracket}
\newcommand{\ssaphi}[1]{{\mathbb F}_{\tt ssa}\llbracket#1\rrbracket}
\newcommand{\ssaassign}[1]{{\mathbb A}_{\tt ssa}\llbracket#1\rrbracket}
\newcommand{\ssaassigns}[1]{{\overline{\mathbb A}}_{\tt ssa}\llbracket#1\rrbracket}
\newcommand{\mayfail}[1]{{#1}_{\tt ex}}

\newcommand{\fjlam}{FJ$_{\lambda}$}

\newcommand{\tarmeth}[1]{{\mathbb M\mathbb D}_{\tt fj\lambda}\llbracket#1\rrbracket}
\newcommand{\tarexp}[1]{{\mathbb E}_{\tt fj\lambda}\llbracket#1\rrbracket}
\newcommand{\tarstmt}[1]{{\mathbb S}_{\tt fj\lambda}\llbracket#1\rrbracket}
\newcommand{\tarvdecl}[1]{{\mathbb V\mathbb D}_{\tt fj\lambda}\llbracket#1\rrbracket}
\newcommand{\tarassign}[1]{{\mathbb A}_{\tt fj\lambda}\llbracket#1\rrbracket}

\newcommand{\cpsconvmeth}[1]{{\mathbb C}{\mathbb M\mathbb D}_{\tt cps}\llbracket#1\rrbracket}
\newcommand{\cpsconvexp}[1]{{\mathbb C}{\mathbb E}_{\tt cps}\llbracket#1\rrbracket}

\newcommand{\cpsconvvdecl}[1]{{\mathbb C}{\mathbb V\mathbb D}_{\tt cps}\llbracket#1\rrbracket}
\newcommand{\cpsconvblk}[1]{{\mathbb C}{\mathbb B}_{\tt cps}\llbracket#1\rrbracket}

\newcommand{\cpsconvphi}[1]{{\mathbb C}{\mathbb F}_{\tt cps}\llbracket#1\rrbracket}
\newcommand{\cpsconvjump}[1]{{\mathbb C}{\mathbb K}_{\tt cps}\llbracket#1\rrbracket}
\newcommand{\cpsconvassign}[1]{{\mathbb C}{\mathbb A}_{\tt cps}\llbracket#1\rrbracket}

\newcommand{\alllambda}{{\Lambda}}
\newcommand{\flowf}[1]{\llbracket #1 \rrbracket}
\newcommand{\flowftwo}[2]{\llbracket #1 \rrbracket({#2})}
\newcommand{\flowfthree}[3]{\llbracket #1 \rrbracket({#2})({#3})}
\newcommand{\retstmt}[1]{{\tt returnStmt}(#1)}
\newcommand{\caller}[1]{{\tt caller}(#1)}
\newcommand{\eval}[2]{{\tt eval}(#1, #2)}
\newcommand{\farg}[1]{{\tt formalArgs}(#1)}
\newcommand{\unrec}{{\tt unreachable}}

\newtheorem{ex}{Example}
        {\end{ex}}

\newenvironment{ttprog}{\begin{trivlist}\small\item \tt
        \begin{tabbing}}{\end{tabbing}\end{trivlist}}

\newtheorem{definition}{Definition}

\newtheorem{eg}{Example}

\newenvironment{example}{
        \begin{eg}\rm\normalfont}%
    {\hfill$\Box$\end{eg}}

\begin{document}

\title{Control Flow Obfuscation for FJ using
  Continuation Passing}         
\subtitle{Extended Version}                     
\author{Kenny Zhuo Ming Lu}
\affiliation{
  \department{School of Information Technology}              
  \institution{Nanyang Polytechnic}            
}
\email{luzhuomi@gmail.com}         
 \affiliation{
  \department{Information Systems Technology and Design}              
  \institution{Singapore University of Technology and Design}            
}
\email{kenny\_lu@sutd.edu.sg}         

\begin{abstract}

  Control flow obfuscation deters software reverse engineering attempts
  by altering the program's control flow transfer. The alternation should not
  affect the software's run-time behaviour. In this paper, we propose a control flow obfuscation approach for
  FJ with exception handling.
  The approach is based on a source to source transformation
  using continuation passing style (CPS).
  We argue that the proposed CPS transformation causes malicious attacks
  using context insensitive static analysis and context sensitive
  analysis with fixed call string to lose precision. 
\end{abstract}



\maketitle

\section{Introduction} \label{sec:intro}


Java applications are ubiquitous thanks to the wide adoption of android
devices. Since Java byte-codes are close to their source codes, it is easy
to decompile Java byte-codes back to source codes with tools.
For example, {\tt javap} shipped with JVM \cite{oracle-java} can be
used to decompile Java class files back to Java source.
This makes the Man-At-The-End attack as one of the major
security threat to Java applications. Code obfuscation is one of the
effective mechanism to deter malicious attack through decompilation. 
There are many obfuscation techniques operating on the level of byte-codes \cite{CHAN20041,
  DBLP:journals/corr/abs-1901-04942, BalachandranCS16}. 
In the domain of source code obfuscation, we find solutions such as \cite{proguard}
applying layout obfuscation. We put our interest in control flow
obfuscation techniques, which include control flow flattening
\cite{Laszlo_obfuscatingc++, Wang:2000:STR:900898, Cappaert:2010:GMH:1866870.1866877} and
continuation passing \cite{DBLP:conf/pepm/Lu19}.
Note that the difference between bytecode obfuscation and source code obfuscation is
insignificant, because of the strong correlation between the Java
bytecodes and source codes. In this paper, we
propose an extension to the continuation passing approach to obfuscate FJ
with exception handling.

We assume the attackers gain access to the byte-codes to which layout
obfuscation has been applied. The attackers decompile the byte-codes
into source codes and attempt to extract secret information by running
control flow analysis on the decompiled code. Our goal here is to cause
the control flow analysis become imprecise or more costly in computation.

\section{Motivating Example} \label{sec:example}
\begin{example} \label{ex:fib-ext}
To motivate the main idea, let's consider the following Java code snippet \\
\begin{lstlisting}
class FibGen {
  int f1, f2, lpos;
  FibGen() {
    f1 = 0; f2 = 1; lpos = 1;
  }
  int get(int x) {
    int i = lpos;                   // (1)
    int r = -1;
    try {                           // (2)
      if (x < i) {                  // (3)
        throw new Exception();      // (4)
      } else {                      
        while (i < x) {             // (5)
          int t = f1 + f2;          // (6)
          f1 = f2; f2 = t; i++;
        }    
      }
      lpos = i;                     // (7)
      r = f2;
    } catch (Exception e) {         // (8)
      println("the input should be greater than " + i + ".");
    }
    return r;                       // (9)
  }
}
\end{lstlisting}
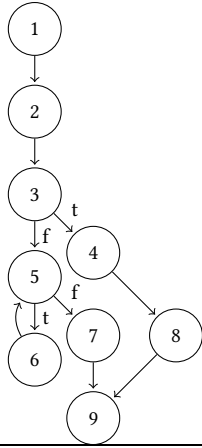
\figtwocol{f:cfg}{CFG of {\tt get}} {
\center
{\footnotesize
\begin{tikzpicture}[shorten >=1pt,node distance=1.1cm,on grid,auto] 
   \node[state] (l1)  {1}; 
   \node[state] (l2) [below =of l1] {2};
   \node[state] (l3) [below =of l2] {3};
   \node[state] (l4) [below right =of l3] {4};   
   \node[state] (l5) [below =of l3] {5};
   \node[state] (l6) [below =of l5] {6};
   \node[state] (l7) [below right =of l5] {7};
   \node[state] (l8) [right =of l7] {8};            
   \node[state] (l9) [below =of l7] {9};
   \path[->]
   (l1) edge  node {} (l2)
   (l2) edge  node {} (l3)
   (l3) edge  node {t} (l4)
   (l3) edge  node {f} (l5)
   (l5) edge  node {t} (l6)
   (l5) edge  node {f} (l7)            
   (l7) edge  node {} (l9)
   (l6) edge[bend left]  node {} (l5)
   (l4) edge  node {} (l8)
   (l8) edge  node {} (l9)                  
   ;
 \end{tikzpicture}     
}
}
In the above we define a Fibonacci number generator in class {\tt FibGen}.
In the method {\tt get}, we compute the Fibonacci number
given the position as the input. Note that the generator maintains a state, in which
we record the last two computed Fibonacci numbers, namely, {\tt f1}
and {\tt f2} and the last computed position 
{\tt lpos}.  In method {\tt get} lines 10 and 11, we raise an exception if the
given input is smaller than {\tt i} which has been initialized to {\tt
  lpos}. Towards the end of the method, we catch the exception and print out the
error message. 

The number comment on the right of each statement indicates the
code block to which the statement belongs.  In Figure~\ref{f:cfg}, we
represent the function {\tt get}'s control flow as a graph. 
Each circle denotes a code block from the source program. 
\end{example}
Inspired by the approach \cite{DBLP:conf/pepm/Lu19}, our main idea
is to translate control flow constructs, such as sequence, if-else, loop into CPS
combinators. In the context of FJ with exception
handling, we translate try-catch statements into CPS combinators as well.

In Figure~\ref{f:get_cps_flatten} we find the obfuscated code snippet of {\tt
  get} method in CPS style. The obfuscated code is in a variant of FJ, named
\fjlam, which is FJ with higher order functions, nested function
declaration and mutable variables in function closures. 
 {\tt  void => void} denotes a function type whose values accept no argument and return
no result. {\tt Exception => void} denotes a function type that
accepts an exception and returns no result. {\tt type NmCont = void =>
  void} defines a type alias. {\tt (void n) -> \{return i < x\};}
defines an anonymous function whose input is of type {\tt void} and
the body returns a boolean value. For brevity, we omit the type annotations
of the formal arguments where there is no confusion. The {\tt return}
key word is omitted when there is only one statement in the function
body. We omit curly brackets in curried expressions, e.g. {\tt x ->
  raise -> k -> \{ ... \}} is the same as {\tt x -> \{ raise -> \{ k
  -> \{ ... \} \} \}}, where {\tt x}, {\tt raise} and {\tt k} are
formal arguments for the lambda abstractions.
For convenience, we treat
method declaration and lambda declaration as
interchangeable. For instance, the lambda declaration
\lstset
{ 
    language=java,
    basicstyle=\footnotesize,
    numbers=none,
    stepnumber=1,
    showstringspaces=false,
    tabsize=1,
    breaklines=true,
    breakatwhitespace=false,
}
\begin{lstlisting}
int => int => int f = x -> y -> { x + y }
\end{lstlisting}
is equivalent to the following method declaration
\begin{lstlisting}
int f (int x, int y) { return x + y;}
\end{lstlisting}

In the last section, we mentioned that the layout obfuscation such as identifier
renaming should have been applied to the obfuscated code; however in this paper we keep all the
identifiers in the obfuscated code unchanged for the ease of
reasoning. For the sake of assessing the obfuscation potency, we
``flatten'' the nested function calls into sequences of assignment
statements. For example, let {\tt x} and {\tt y} be variables of type
{\tt int}, let {\tt f} be a function of type {\tt int =>
  int => int} and {\tt g} be a function of type {\tt int => int}; 
instead of {\tt int r = f(x)(g(y)); }, we write:
\begin{lstlisting}
  int => int f_x = f(x);
  int g_y = g(y);
  int r = f_x(g_y);
\end{lstlisting}  
As we  observe in Figure~\ref{f:get_cps_flatten}, all the
building blocks are continuation functions with type {\tt CpsFunc}.
The simple code blocks  (1), (6), (7) and (8) from the original source
code, which contain no control flow branching statements, are translated into
nested CPS functions {\tt get1}, {\tt get6}, {\tt get7} and {\tt
  get8}. Block (4) containing a {\tt throw} statement is translated
into {\tt get4} which applies the exception object to the exception
handling continuation {\tt raise} of type {\tt ExCont}. Block (9)
has a {\tt return} statement, which is translated into a function in
which we assign the variable being returned {\tt r} to the {\tt res}
variable and call the normal continuation {\tt k} of type {\tt NmCont}. Block (2) is a try
catch statement which is encoded as a call to the {\tt trycatch}
combinator in line 24. Similarly block (3) the if-else statement is
encoded as a call to the {\tt ifelse} combinator and block (5) the
while loop is encoded as a call to the {\tt loop} combinator. 

In Figure~\ref{f:combinators_flatten},
we present the definitions of the CPS combinators used in the
obfuscation.
Combinator {\tt loop} accepts a condition test {\tt cond}, a
continuation executor {\tt visitor} to be executed when the condition
is satisfied, a continuation executor {\tt exit} to be activated when the
condition is not satisfied.
Combinator {\tt seq} takes two continuation
executors and executes them in sequence.
Combinator {\tt trycatch} takes a continuation executor {\tt tr} and
an exception handling continuation {\tt hdl}. It executes {\tt tr} by
replacing the current exception continuation with {\tt ex\_hdl}. 
Combinator {\tt ifelse} accepts a condition test {\tt cond}, a
continuation for the then-branch {\tt th} to be executed when the condition
is satisfied, a continuation executor for the else-branch {\tt el} to be activated when the
condition is not satisfied.

\lstset
{ 
    language=java,
    basicstyle=\footnotesize,
    numbers=none,
    stepnumber=1,
    showstringspaces=false,
    tabsize=1,
    breaklines=true,
    breakatwhitespace=false,
}

\begin{figure}
\begin{lstlisting}
type ExCont = Exception => void;
type NmCont = void => void; 
type CpsFunc = ExCont => NmCont => void;
int get(int x) {
  int i, t, r, res; Exception ex;
  int => ExCont => (int => void) => void get_cps = 
  x -> raise -> k -> {
    void => bool cond5 = n-> { i < x}; 
    void => bool cond3 = n -> {x < i}; 
    CpsFunc get5 = loop(cond5, get6, get7)
    CpsFunc get3 = ifelse(cond3,get4,get5);
    CpsFunc get1_2 = seq(get2, get9);
    CpsFunc pseq = seq(get1, get1_2);
    NmCont => void pseq_raise = pseq(raise);
    NmCont nk_res = n->k(res);
    return pseq_raise(nk_res);
  }
  CpsFunc get1 = (ExCont raise) -> (NmCont k) -> {
    i = this.lpos; r = -1; return k();
  }

  Exception => CpsFunc hdl =
    e -> {ex = e; return get8;}
  CpsFunc get2 = trycatch( get3, hdl);

  CpsFunc get4 = raise -> k
    -> raise(new Exception());
  CpsFunc get6 = raise -> k
    -> { t = this.f1 + this.f2; this.f1 = this.f2;
         this.f2 = t; i = i + 1; return k();}
  CpsFunc get7 = raise -> k 
    -> { this.lpos = i; r = this.f2; return k();}
  CpsFunc get8 = raise -> k 
    -> { System.out.println("..."); return k();}
  CpsFunc get9 = raise -> k
    -> { res = r; return k(); }
  NmCont id_bind = i -> { res = i; return; };
  CpsFunc get_x = get_cps(x);     
  NmCont => void get_x_raise = get_x(id_raise);
  void ign = get_x_raise(id_bind);
  return res;
}
void id_raise(Exception e) { return ;}
\end{lstlisting}

\hrule
\vspace{-3mm}
\caption{{\tt get} in CPS (flatten) (Line 1-43)} \label{f:get_cps_flatten}
\end{figure}

\begin{figure}
\begin{lstlisting}[firstnumber=50]
CpsFunc loop(void => Boolean cond, 
  CpsFunc visitor, CpsFunc exit) {
  return raise -> k -> {
    if (cond()) { 
      NmCont => void visitor_raise = visitor(raise);
      NmCont nloop = n -> { 
        CpsFunc ploop = loop(cond, visitor, exit);
        NmCont => void = ploop_raise = ploop(raise);
        return ploop_raise(k);
      };
      return visitor_raise(nloop);
    } else {
      NmCont => void exit_raise = exit(raise);
      return exit_raise(k);
    }
  }
}
CpsFunc seq(CpsFunc first, CpsFunc second) {
  return raise -> k -> {
    NmCont => void first_raise = first(raise);
    NmCont n_second = n -> { 
      NmCont => void second_raise = second(raise);
      return second_raise(k);
    };
    return first_raise(n_second);
  }
}
CpsFunc trycatch(CpsFunc tr, Exception => CpsFunc hdl) {
  return raise -> k -> {
    ExCont ex_hdl = ex -> { 
      CpsFunc hdl_ex = hdl(ex);
      NmCont => void hdl_ex_raise = hdl_ex(raise);
      return hdl_ex_raise(k);
    }
    NmCont => void tr_hdl = tr(ex_hdl);
    return tr_hdl(k);
  }
}
CpsFunc ifelse(void => Boolean cond,
  CpsFunc th, CpsFunc el) {
  return raise -> k -> {
    if (cond()) {
      NmCont => void th_raise = th(raise); 
      return th_raise(k); 
    } else {
      NmCont => void el_raise = el(raise); 
      return el_raise(k); 
    }
  }
}
\end{lstlisting}
\hrule
\vspace{-3mm}
\caption{CPS Combinators (flatten) (Line 50-99)} \label{f:combinators_flatten}
\end{figure}

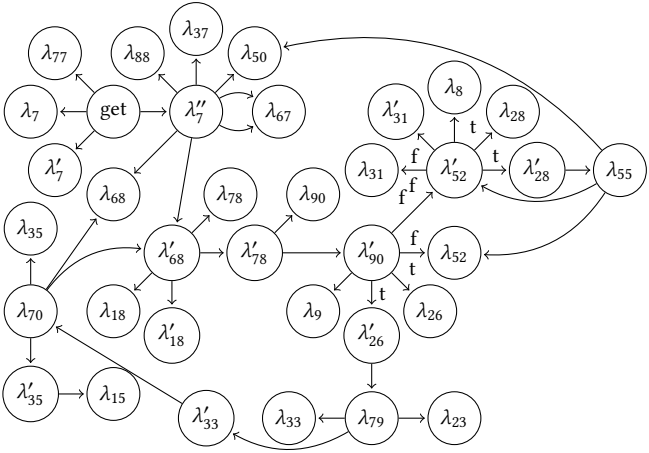
\figtwocol{f:cfg_obfs}{Reconstructed CFG of {\tt get}} {
{\footnotesize

\begin{tikzpicture}[shorten >=1pt,node distance=1.1cm,on grid,auto] 
   \node[state] (get)   {get}; 
   \node[state] (lam77) [above left=of get] {$\lambda_{77}$};
   \node[state] (lam7) [left=of get] {$\lambda_{7}$};
   \node[state] (lam7p) [below left=of get] {$\lambda_{7}'$};
   \node[state] (lam68) [below =of get] {$\lambda_{68}$};
   \node[state] (lam7pp) [right =of get] {$\lambda_7''$};
   \node[state] (lam88) [above left =of lam7pp] {$\lambda_{88}$};
   \node[state] (lam37) [above =of lam7pp] {$\lambda_{37}$};
   \node[state] (lam50) [above right =of lam7pp] {$\lambda_{50}$};
   \node[state] (lam67) [right =of lam7pp] {$\lambda_{67}$};
   \node[state] (lam68p) [below right =of lam68] {$\lambda_{68}'$};
   \node[state] (lam18) [below left =of lam68p] {$\lambda_{18}$};
   \node[state] (lam18p) [below =of lam68p] {$\lambda_{18}'$};
   \node[state] (lam78) [above right =of lam68p] {$\lambda_{78}$};
   \node[state] (lam78p) [right =of lam68p] {$\lambda_{78}'$};
   \node[state] (lam90) [above right =of lam78p] {$\lambda_{90}$};
   \node[state] (lam90p) [below right =of lam90] {$\lambda_{90}'$};
   \node[state] (lam52) [right =of lam90p] {$\lambda_{52}$};
   \node[state] (lam52p) [above =of lam52] {$\lambda_{52}'$};
   \node[state] (lam26) [below right =of lam90p] {$\lambda_{26}$};
   \node[state] (lam26p) [below =of lam90p] {$\lambda_{26}'$};
   \node[state] (lam9) [below left =of lam90p] {$\lambda_{9}$};
   \node[state] (lam31) [left =of lam52p] {$\lambda_{31}$};
   \node[state] (lam31p) [above left =of lam52p] {$\lambda_{31}'$};
   \node[state] (lam8) [above =of lam52p] {$\lambda_{8}$};
   \node[state] (lam28) [above right =of lam52p] {$\lambda_{28}$};
   \node[state] (lam28p) [right =of lam52p] {$\lambda_{28}'$};
   \node[state] (lam55) [right =of lam28p] {$\lambda_{55}$};
   \node[state] (lam79) [below =of lam26p] {$\lambda_{79}$};
   \node[state] (lam23) [right =of lam79] {$\lambda_{23}$};
   \node[state] (lam33) [left =of lam79] {$\lambda_{33}$};
   \node[state] (lam33p) [left =of lam33] {$\lambda_{33}'$};
   \node[state] (lam70) [left =of lam18] {$\lambda_{70}$};         
   \node[state] (lam35p) [below =of lam70] {$\lambda_{35}'$};   
   \node[state] (lam35) [above =of lam70] {$\lambda_{35}$};
   \node[state] (lam15) [right =of lam35p] {$\lambda_{15}$};
    \path[->] 
    (get) edge  node {} (lam77)
    (get) edge  node {} (lam7)
    (get) edge  node {} (lam7p)
    (get) edge  node {} (lam7pp)
    (lam7pp) edge node {} (lam68)
    (lam7pp) edge node {} (lam88)
    (lam7pp) edge node {} (lam37)
    (lam7pp) edge node {} (lam50)
    (lam7pp) edge[bend left] node {} (lam67)                
    (lam7pp) edge[bend right] node {} (lam67)                
    (lam7pp) edge node {} (lam68p)                    
    (lam68p) edge node {} (lam18) 
    (lam68p) edge node {} (lam18p)
    (lam68p) edge node {} (lam78) 
    (lam68p) edge node {} (lam78p)
    (lam78p) edge node {} (lam90)
    (lam78p) edge node {} (lam90p)
    (lam90p) edge node {f} (lam52p)
    (lam90p) edge node {f} (lam52)    
    (lam90p) edge node {t} (lam26p)
    (lam90p) edge node {t} (lam26)
    (lam90p) edge node {} (lam9)        
    (lam52p) edge node {f} (lam31)
    (lam52p) edge node {f} (lam31p)
    (lam52p) edge node {} (lam8)            
    (lam52p) edge node {t} (lam28)
    (lam52p) edge node {t} (lam28p)
    (lam28p) edge node {} (lam55)
    (lam55) edge[bend left] node {} (lam52)
    (lam55) edge[bend left] node {} (lam52p)
    (lam55) edge[bend right] node {} (lam50)
    (lam26p) edge node {} (lam79)
    (lam79) edge node {} (lam23)
    (lam79) edge node {} (lam33)
    (lam79) edge[bend left]  node {} (lam33p)
    (lam33p) edge  node {} (lam70)
    (lam70) edge[bend left] node {} (lam68p)
    (lam70) edge node {} (lam68)
    (lam70) edge node {} (lam35p)
    (lam70) edge node {} (lam35)
    (lam35p) edge node {} (lam15)    
    ;
\end{tikzpicture}     
}
}

To assess the potency of the obfuscation technique, let's put on the hat
of the attackers and apply some static analysis to the obfuscated
source code. The goal of the attack is to reconstruct the control flow
graph from the obfuscated source. We apply an inter-procedural data
flow analysis to the obfuscated code. For each variable or formal
argument in the code, the analysis tries to
approximate the set of possible lambda expressions which the
variable/argument may capture during the execution. From the
approximation we re-create the (global) control flow graph as
presented in Figure~\ref{f:cfg_obfs}. We give names to anonymous functions as $\lambda_l$ where $l$
refers to the line number appearing in Figures~\ref{f:get_cps_flatten}
and \ref{f:combinators_flatten}.
In case that there are more than one anonymous functions introduced in line $l$. We use
$\lambda_i$ to denote the first one, $\lambda_i'$ to denote the second one and $\lambda_i''$ to denote the third one.
Compared to the CFG of the original source, the reconstructed CFG of the obfuscated code are far more complex.
For instance, there exist more than one loop in the obfuscated CFG in Figure~\ref{f:cfg_obfs}, namely,
\begin{enumerate}
\item $\lambda_{52}', \lambda_{28}', \lambda_{55}, \lambda_{52}'$,
\item $\lambda_{68}', \lambda_{78}', \lambda_{90}', \lambda_{26}', \lambda_{79}, \lambda_{33}', \lambda_{70}, \lambda_{68}'$,
\end{enumerate}
whereas there is clearly only one loop in the original CFG in Figure~\ref{f:cfg}.
The loss of precision is due to the fact that the attack which we
simulate is using a context insensitive data flow analysis, which is
known to be incomplete in
the presence of multiple calls to the same function. For example, in
Figure~\ref{f:get_cps_flatten}, lines 12 and 13, we call the
combinator {\tt seq} twice with different actual arguments. The
analysis ignores the context and union the two sets of actual
arguments into sets. These approximation are propagated along to the
rest of the analysis. A similar observation is applicable to context
sensitive analysis with a fixed size call string, in the
presence of multiple calls to recursive combinators such as {\tt loop}.
Attackers may choose to use a context
sensitive analysis, however to achieve a better approximation, the
analysis will be much more costly and often not practical.

For the ease of establishing correctness result, we use an extension
of the Single Static Assignment form for FJ with exception handling
(SSAFJ-EH) as the source language of the translation.
The construction of SSAFJ-EH can be extended from the work found in 
the literature \cite{DBLP:conf/ecoop/AnconaC16}, which is not the focus
of this paper, hence we omit the details.

The contributions of this paper include,
\begin{itemize}
  \item We formalize the single static assignment form of FJ with Exception Handling.
  \item We develop a control flow obfuscation algorithm by
    translating SSAFJ-EH to \fjlam using continuation passing.
  \item We show that CPS based control flow obfuscation is effective
    against static analysis, in particular context insensitive control
    flow analysis.
\ignore{    
  \item We implement the obfuscation strategy in a Java
    obfuscater. We conduct benchmarking experiments to show that our
    approach is practical.
}    
\end{itemize}

The rest of the paper is organized as follows,
In Section ~\ref{sec:ssa}, we formalize SSAFJ-EH's syntax and semantics. 
In Section~\ref{sec:translation}, we define the syntax of \fjlam as well as its semantics.
We formalize the source-to-source translation from
SSAFJ-EH to \fjlam. 
In Section~\ref{sec:analysis}, we discuss in details about the potency
assessment of our obfuscation technique against static analyses. 
We discuss about related works in 
Section~\ref{sec:related-works}  and conclude in Section~\ref{sec:conclusion}.

\section{Single Static Assignment Form for FJ with Exception Handling} \label{sec:ssa}
\subsection{Syntax of SSAFJ-EH} \label{sec:ssa_syntax}

We extend the syntax of SSAFJ \cite{DBLP:conf/ecoop/AnconaC16} with exception handling,

{\footnotesize
  \bda{rcl}
  \tlabel{ClassDecl} ~~~~ \cd & ::= & \CLASS~C~\{\overline{\fd}; \overline{\md}\} \\
  \tlabel{FieldDecl} ~~~~ \fd & ::= & t~f \\
  \tlabel{MethodDecl} ~~~~ \md & ::= & t~m~(t~x)~\{\overline{\vd}; \overline{b} \} \\
  \tlabel{VarDecl} ~~~~ \vd & ::= & t~x \\ 
  \tlabel{Block} ~~~~ b & ::= & l : \{ s \} \\
  \tlabel{Statement} ~~~~ s & ::= & \overline{a} \mid \RET~e \mid \THROW~e \mid x = e.m(e) 
  \\ & &  \mid \TRY \{\overline{b}\}~\JOIN~\{\overline{\phi}\}~\CATCH~(t~x)~\{ \overline{b} \}~\JOIN~\{\overline{\phi}\}
  \\ & & \mid \JOIN~\{\overline{\phi}\}~ \WHILE~e~\{\overline{b} \}
  \\ & & \mid \IF~ e~ \{ \overline{b} \}~ \ELSE~ \{\overline{b} \}~\JOIN~\{\overline{\phi}\}
  \\ 
  \tlabel{Assignment} ~~~~ a & ::= & x~ =~ e | e.f~ =~ e \\
  \tlabel{Phi} ~~~~ \phi & ::=  & x = \PHI(\overline{l:x}) \\ 
  \tlabel{Label} ~~~~ l & ::= & L_0 \mid L_1 \mid L_2 \mid ... \\ 
  \tlabel{Expression} ~~~~ e  & ::= & v \mid x \mid e.f \mid \NEW~t() \mid \THIS \mid e ~op~ e \\
  \tlabel{Operator} ~~~~ op & ::= & + \mid - \mid  > \mid < \mid == \mid ... \\
  \tlabel{Type} ~~~~ t & ::= & int \mid bool \mid void \mid C \\ 
  \tlabel{Value} ~~~~ v & ::= & c \mid loc \mid \NULL \\
  \tlabel{MemLoc} ~~~~ loc & ::= & \loc{0} \mid \loc{1} \mid ... 
\eda
}
$\CLASS~C~\{\overline{\fd}; \overline{\md}\}$ defines a class. 
$C$ denotes a class name. $\overline{\fd}$ denotes a sequence of field declarations, $\fd_1; ... ; \fd_n$.
Likewise for $\overline{\md}$ denotes a sequence of method
declarations. 
For simplicity, we do not consider class inheritance, class constructors and method modifiers. 
Implicitly we assume each class comes with a default constructor and all field declarations are public and non-static.
$t~m~(t~x)~\{\overline{\vd}; \overline{b} \}$ defines a method declaration. For
simplicity, we restrict the language to single
argument methods. $m$ denotes a method name. $x$, $y$ and $z$ denote variables.
$\overline{b}$ defines a sequence of blocks. Each block
is associated with a label $l$. Labels are unique
within the method body. Reference to labels is restricted to the
method's local scope. Each block consists of a sequence of assignment statements or a control flow statement.
The last block in a method must contain a return statement.
Note that all control flow statements potentially alter the default
top-down execution order. The SSA form ensures the definition of a
variable through assignment must dominate all the uses of this variable.
Unlike the work \cite{DBLP:conf/pepm/Lu19}, which uses low-level SSA structure with goto
statements, the SSA form introduced in this paper is in a high-level structured
form. That is, only certain control flow statements,  such as if-else,
try-catch and while,
may carry one or more $\phi$ clauses. There is no goto statement. 
A $\phi$ assignment $x = \PHI(\overline{l:x})$ selects the right labeled
argument $l_i:x_i$ to assign to the left hand side variable, based on the label
of the preceding statement. For if-else statement, the $\phi$
assignment is inserted right after the then- and else-branches, which merges
the possible different set of values from the branches into a new set of
variables. In the while loop, the $\phi$ assignment is located before
the loop-condition. In try-catch statement, we find two sets of $\phi$
assignments. The $\phi$ assignments located after the try block and catch block
has a functionality similar to the one in if-else statement. The other
one is located between the try block and the catch block. It is to merge
the different sets of values that are arising in various parts of the
try block due to exception being raised. We will discuss more in
details in the semantics of SSAFJ-EH. $t$ denotes a type.
A type $t$ can be basic types such as $int$, $void$ or a class type $C$.
A value $v$ is either a constant, a memory location or $\NULL$.
The formal details will be elaborated
in the upcoming subsection. Syntax of assignments and
expressions is standard.  For instance, the corresponding SSA form of the method {\tt
  get} from the class {\tt FibGen} in Example~\ref{ex:fib-ext} is in Figure~\ref{f:get_ssa}.

\begin{figure} 
  \begin{lstlisting}
int get(int x) {
    int i_1, i_2, i_5, i_6, t_6, r_1, r_2, r_7;
L1: i_1 = this.lpos;
    r_1 = -1;
L2: try {
    L3: if (x < i_1) {
        L4: throw new Exception();
        } else { 
        L5: join {i_5=phi(L3:i_1, L6:i_6)} while (i_5 < x) {
            L6: t_6 = this.f1 + this.f2;
                this.f1 = this.f2;
                this.f2 = t_6;
                i_6 = i_5 + 1;
            }
        L7: this.lpos = i_5;
            r_7 = this.f2;
        } join {i_3 = phi(L4:i_1, L7:i_5)}
    } join {i_2 = phi(L4:i_1)}
      catch (Exception e) {
    L8: System.out.println("the input..." + i_2 + ".");
    } join {r_2 = phi(L3:r_7, L8:r_1)};
L9: return r_2; 
}
  \end{lstlisting}
\hrule
\vspace{-3mm}
\caption{Method {\tt get} in Single Static Assignment Form}  \label{f:get_ssa}
\end{figure}

\subsection{Semantics of SSAFJ-EH}  \label{sec:ssa_semantics}

\figtwocol{f:ssa_semantics_p1}{Denotational Semantics of SSAFJ-EH (Part 1)}{
{\footnotesize
  \bda{rrcl}
  \tlabel{Global Decl Env} & {\tt GEnv} & \subseteq & ({\tt Class} \times \overline{\tt FieldDecl}) ~ \cup \\
  & & &  (({\tt Class} \times {\tt MethodName}) \times {\tt MethodDecl}) \\
  \tlabel{Local Decl Env} & {\tt LEnv} & \subseteq & ({\tt Variable} \times {\tt Value}) \\
  \tlabel{Memory Store} & {\tt Store} & \subseteq & ({\tt MemLoc} \times {\tt Object}) \\
  \tlabel{Object} & \OBJ & ::= & \OBJ(t,\rho)  \\
  \tlabel{Exception} & ex & ::= & \EXCEPTION(v,{\tt LEnv}, {\tt Store}, l) \\   
  \tlabel{Object Field Map} & \rho & \subseteq & ({\tt FieldName} \times {\tt Value})
  \eda

  \bda{ll}
    \multicolumn{2}{l}{\ssameth{\cdot} :: {\tt MethodDecl} \arrow {\tt
     Value} \arrow {\tt Value} \arrow {\tt GEnv} \arrow {\tt Store} \arrow } \\
    \multicolumn{2}{l}{\bgap\bgap\bgap\bgap\bgap\bgap \mayfail{({\tt Value}, {\tt Store})}
    }\\ 

    \ssameth{(t'~ m(t~x)\{\overline{\vd}; \overline{b}\})}\ v_o\ v_x\ genv\ st
    =
    \\ \bgap \LET~lenv'= \ssavdecl{\overline{\vd}}~\{(this,v_o), (x,v_x)\}
    \\ \bgap \IN~\CASE~\ssablks{\overline{b}}~L_0~genv~lenv'~st~ \OF
    \\ \bgap \bgap \EXCEPTION(v, lenv'', st'. l') \arrow \EXCEPTION(v, lenv', st',L_0) 
    \\ \bgap \bgap (v, lenv'', st', l') \arrow (v, st') 
  \eda
    
 \bda{ll}
 \multicolumn{2}{l}{\ssablks{\cdot} :: [{\tt Block}] \arrow {\tt Label} \arrow {\tt GEnv} \arrow {\tt LEnv} \arrow {\tt Store} \arrow } \\
 \multicolumn{2}{l}{ \bgap\bgap\bgap\bgap\bgap\bgap  \mayfail{({\tt Value}, {\tt LEnv}, {\tt Store}, {\tt Label})}}\\ 
 \ssablks{b}~l~genv~lenv~st~ = & \ssablk{b}~l~genv~lenv~st \\
 \ssablks{b;\overline{b}}~l~genv~lenv~st~ = & \CASE~\ssablk{b}~l~genv~lenv~st~\OF \\
 \multicolumn{2}{l}{ \bgap\bgap\bgap\bgap \EXCEPTION(v,lenv',st',l') \arrow \EXCEPTION(v, lenv',st', l)}\\ 
 \multicolumn{2}{l}{ \bgap\bgap\bgap\bgap (v,lenv',st',l') \arrow \ssablks{\overline{b}}~l'~genv~lenv'~st'} 
  \eda
  \bda{l}

\ssablk{\cdot} :: {\tt Block} \arrow {\tt Label} \arrow {\tt
  GEnv} \arrow {\tt LEnv}  \arrow {\tt Store} \arrow \\
\bgap \bgap \bgap \bgap \mayfail{({\tt Value}, {\tt LEnv}, {\tt Store}, {\tt Label})} \\  
 \ssablk{l:\{\IF~(e)~\{\overline{b_1}\}~\ELSE~\{\overline{b_2}\}~\JOIN\{\overline{\phi} \} \}}~l_p~genv~lenv~st = 
 \\ 
 \bgap \bgap  \CASE~\ssaexp\{e\}~genv~lenv~st~\OF \\
 \bgap \bgap  (true, st') \arrow \CASE~\ssablks{\overline{b_1}}~l~genv~lenv~st'~\OF \\
 \bgap \bgap \bgap \bgap \EXCEPTION(v,lenv',st'',l') \arrow \EXCEPTION(v,lenv', st'',l') \\
 \bgap \bgap \bgap \bgap(v,lenv',st'',l') \arrow (v, \ssaphi{\overline{\phi}}~l'~lenv', st'', l) \\
 \bgap \bgap  (false, st') \arrow \CASE~\ssablks{\overline{b_2}}~l~genv~lenv~st'~\OF \\
 \bgap \bgap \bgap \bgap \EXCEPTION(v,lenv',st'',l') \arrow \EXCEPTION(v,lenv', st'',l') \\
 \bgap \bgap \bgap \bgap(v,lenv',st'',l') \arrow (v, \ssaphi{\overline{\phi}}~l'~lenv', st'', l) \\
 
 \ssablk{l:\{ \RET~e \}}~l_p~genv~lenv~st = ~\CASE~\ssaexp{e}~genv~lenv~st~\OF\\
 \bgap \bgap \bgap \bgap (v,st') \arrow (v, lenv, st', l)  \\
 
 \ssablk{l:\{ \THROW~e \}}~l_p~genv~lenv~st = ~\CASE~\ssaexp{e}~genv~lenv~st~\OF\\
 \bgap \bgap \bgap \bgap (v,st') \arrow \EXCEPTION(v, lenv, st', l)  \\
 
 \ssablk{l:\{ \TRY \{ \overline{b} \} ~\JOIN \{\overline{\phi_r}\}~\CATCH(t~x)~\{\overline{b'}\}~\JOIN~\{\overline{\phi_k}\}}~l_p~genv~lenv~st = \\
 \bgap \CASE~\ssablks{\{\overline{b}\}}~genv~lenv~st~\OF \\
 \bgap (v,lenv',st',l') \arrow (v, \ssaphi{\overline{\phi_k}}~l'~lenv', st', l') \\ 
 \bgap \EXCEPTION(v,lenv',st',l') \arrow \\
 \bgap \bgap \LET~lenv'' = \ssaphi{\overline{\phi_r}}~l'~lenv'+(x,v) \\ 
 \bgap \bgap \IN~\CASE~\ssablks{\overline{b'}}~l'~genv~lenv''~st'~\OF \\
 \bgap \bgap \bgap (v', lenv''', st'', l'') \arrow (v', \ssaphi{\overline{\phi_k}}~l''~lenv''', st'', l'') \\ 
 \bgap \bgap \bgap \EXCEPTION(v', lenv''', st'', l'') \arrow \EXCEPTION(v', lenv''', st'', l'') \\

 \ssablk{l:\{\JOIN~\{\overline{\phi}\}~\WHILE(e)~\{\overline{b}\}\}}~l_p~genv~lenv~st = \\
 \bgap \LET~lenv'~ = ~ \ssaphi{ \overline{\phi} }~l_p~lenv \\
 \bgap \IN~ \CASE~ \ssaexp{e}~genv~lenv'~st~\OF \\
 \bgap (false, st') \arrow (null, lenv', st', l) \\
 \bgap (true, st') \arrow \CASE~\ssablks{\overline{b}}~l~genv~lenv'~st' \OF \\
 \bgap \bgap \EXCEPTION(v,lenv'',st'',l') \arrow \EXCEPTION(v,lenv'',st'',l') \\
 \bgap \bgap (v, lenv'', st'', l') \arrow \\
 \bgap \bgap \bgap \ssablk{ l:\{~\JOIN~\{\overline{\phi}\}~\WHILE(e)~\{\overline{b}\}\} }~l'~genv~lenv''~st'' \\

 \ssablk{l:\{ x = e_1.m(e_2) \}}~l_p~genv~lenv~st = \CASE~\ssaexp{e_1}~genv~lenv~st~\OF \\
 \bgap (\loc{n}, st') \arrow \CASE~st'(\loc{n})~\OF \\
 \bgap \bgap \OBJ(t,\rho) \arrow \CASE~genv(t,m)~\OF \\
 \bgap \bgap \bgap \md \arrow \CASE~\ssaexp{e_2}~genv~lenv~st~\OF \\
 \bgap \bgap \bgap \bgap (v'',st'') \arrow \CASE~\ssameth{\md}~\loc{n}~v''~genv~st''~\OF \\
 \bgap \bgap \bgap \bgap \bgap (v''',st''') \arrow (null, lenv+(x,v'''), st''', l) \\
 \bgap \bgap \bgap \bgap \bgap \EXCEPTION(v''',\_,st''',\_) \arrow \EXCEPTION(v''',lenv, st''',l)  \\

 \ssablk{l:\{ \overline{a} \}}~l_p~genv~lenv~st = \CASE~\ssaassign{\overline{a}}~genv~lenv~st~\OF \\
 \bgap \bgap (lenv',st') \arrow (null, lenv',st',l) 
 \eda

}
}

\figtwocol{f:ssa_semantics_p2}{Denotational Semantics of SSAFJ-EH (Part 2) }{
{\footnotesize

  \bda{ll}
 \multicolumn{2}{l}{\ssavdecl{\cdot} :: [{\tt VarDecl}] \arrow {\tt LEnv} \arrow {\tt LEnv}} 
\\ 
  \ssavdecl{[]}\ lenv  = & lenv \\
  \ssavdecl{(t\ x);\overline{\vd}}\ lenv = &
  \ssavdecl{\overline{\vd}}~(lenv + (x,null))  \\
  \eda

 \bda{ll}
 \ssaphi{\cdot} :: [{\tt Phi}] \arrow {\tt Label} \arrow {\tt LEnv}
  \arrow {\tt LEnv} \\
  \ssaphi{[]} ~l~lenv~  =  lenv \\
  \ssaphi{x = \PHI(l_1:x_1, ,..., l_i:x_i,..., l_n:x_n) ; \overline{\phi}}
  ~l_i~lenv = \\
  \bgap \ssaphi{\overline{\phi}} ~l_i~lenv + (x,x_i)
 \eda

 \bda{ll}
 \ssaassigns{\cdot} :: [{\tt Assignment}] \arrow {\tt GEnv} \arrow {\tt LEnv} \arrow {\tt Store} \arrow ({\tt LEnv}, {\tt Store}) \\
 \ssaassigns{[]}~genv~lenv~st = (lenv, st) \\
 \ssaassigns{ a; \overline{a} }~genv~lenv~st = \CASE~\ssaassign{a}~genv~lenv~st~\OF \\
 \bgap (lenv', st') \arrow \ssaassigns{\overline{a}}~genv~lenv' st' 
 \eda

 \bda{ll}
 \ssaassign{\cdot} :: {\tt Assignment} \arrow {\tt GEnv} \arrow {\tt LEnv} \arrow {\tt Store} \arrow ({\tt LEnv}, {\tt Store}) \\
 \ssaassign{x = e} ~genv~lenv~st = \CASE~\ssaexp{e}~genv~lenv~st~\OF \\
 \bgap (v',st') \arrow (lenv + (x,v'), st') \\
 
 \ssaassign{e.f = e'} ~genv~lenv~st = \CASE~\ssaexp{e}~genv~lenv~st~\OF \\
 \bgap (\loc{n}, st') \arrow \CASE~st'(\loc{n})~\OF \\
 \bgap \bgap \OBJ(t,\rho) \arrow \CASE~\ssaexp{e'}~genv~lenv~st'~\OF \\
 \bgap \bgap \bgap (v,st'') \arrow (lenv, st''+ (\loc{n}, \OBJ(t, \rho + (f,v))))
 \eda

  \bda{ll}
\multicolumn{2}{l}{  \ssaexp{\cdot} :: {\tt Expression} \arrow {\tt GEnv} \arrow {\tt
    LEnv} \arrow {\tt Store} \arrow ({\tt Value}, {\tt Store})}
\\ 
  \ssaexp{v}\ genv\ lenv\ st = (v, st) \\
  \ssaexp{x}\ genv\ lenv\ st = (lenv(x), st) \\
  \ssaexp{\THIS}\ genv\ lenv\ st = (lenv(\THIS), st) \\
  \ssaexp{e.f}\ genv\ lenv\ st = \CASE~\ssaexp{e}~genv~lenv~st~\OF \\
  \bgap (\loc{n},st') \arrow \CASE~st'(\loc{n})~\OF \\
  \bgap \bgap \OBJ(t,\rho) \arrow (\rho(f), st') \\
  \ssaexp{\NEW~t()}\ genv\ lenv\ st = \LET~n = maxloc(st) \\
  \bgap \sgap \rho = \{ (f, \NULL) | f \in genv(t) \} \\
  \bgap \IN~ (\loc{n+1}, st + (\loc{n+1}, \OBJ(t,\rho))) \\
  \ssaexp{e_1~op~e_2}\ genv\ lenv\ st =\CASE~\ssaexp{e_1}~genv~lenv~st~\OF \\
  \bgap (v_1, st_1) \arrow \CASE~\ssaexp{e_2}~genv~lenv~st_1~\OF \\
  \bgap \bgap (v_2, st_2) \arrow  (apply(op, v_1,v_2), st_2) \\ 
  \eda
}
}

We report a call-by-value semantics of SSAFJ-EH in
Figures~\ref{f:ssa_semantics_p1} and~\ref{f:ssa_semantics_p2}.
We adopt the standard denotational semantics notation found in
~\cite{kpeterssonSSPL}. {\tt GEnv} denotes a constant global environment which maps
class names to field declarations and class names and method names to method declarations.
 We assume that the given program is free of type errors
and there is no null pointer reference error. 
{\tt LEnv} denotes a local variable environment which maps variables to values. {\tt Store} defines a
memory environment that maps memory locations to objects.

As a convention, we write $m(a)$ to refer to the
object $b$ associated with the key $a$ in a mapping $m$,
i.e. $(a,b) \in m$, given that all keys in $m$ are unique. We use $m +
(a,b)$ to denote an ``update if exists -- insert otherwise'' operation, i.e. $m + (a,b) = \{ (x,y)
\in m | a \neq x \} \cup \{ (a, b) \}$.

In this paper, we are only interested in the obfuscation of methods, hence we omit the semantics for
class declaration and field declaration. 
$\ssameth{\cdot}$ defines the semantics of a method as a function
expecting a reference to the current object, a value as the actual argument,
a global environment and a memory store and returns
a pair of value and memory store as result. Given a domain $D$, we write $\mayfail{D}$ to denote
$D \cup {\tt Exception}$. 
$\ssavdecl{\cdot}$ takes a list of
variable declarations and a local declaration environment 
as inputs then registers each variable in the declaration
environment. Note that we use Haskell's style of let-binding to
introduce temporary variables and case expression for pattern
matching. For breivity we omitted data constructors in the patterns
when there is no ambiguity.


We adopt Haskell's style list syntax. $[]$ denotes an empty list. $x:xs$
denotes a non-empty list where $x$ refers to the head and $xs$ refers
to the tail. We assume there exists an implicit conversion from a sequence $b_1;b_2;...;b_n$
to a list $b_1:b_2:...:b_n:[]$.

$\ssablk{\cdot}$ evaluates a block with respect to the context, i.e. the label of the preceding block,
the local environment and the memory store. As the output, it returns a tuple of four items, namely,
the value of the evaluation, the updated local environment, the updated memory store and the label from the exiting block
if there is no exception occurred, otherwise an exception is returned.
$\ssablks{\cdot}$ evaluates a sequence of blocks by applying $\ssablk{\cdot}$ to each block in order,
and propagates the resulting environments if there is no exception, otherwise the exception is propagated. 

We highlight the a few interesting cases of $\ssablk{\cdot}$.
In case of if-else statement, we evaluate either the then-branch $\overline{b_1}$ or
the else-branch $\overline{b_2}$ depending on the result of the condition expression $e$.
Given the label of the exiting block, either from $\overline{b_1}$ or $\overline{b_2}$, we apply $\ssaphi{\overline{\phi}}$
to update the local environment in the result.
In case of a try-catch statement, we first evaluate the try block. If the evaluation is
successful, we compute the result by updating the local environment with $\ssaphi{\overline{\phi_k}}$.
If some exception arises from the evaluation of the try block, we generate a local environment with $\ssaphi{\overline{\phi_r}}$ depending on the location from which the exception is raised. Next we evaluate the catch block under this local environment.
Finally we update the output local environment with $\ssaphi{\overline{\phi_k}}$.
In case of a method invocation, we evaluate the object expression into a memory location, from which we look up the memory store
to retrieve the actual object and its type. From the global environment, we retrieve the method declaration based on the
method name $m$. We call $\ssameth{\md}$ with the actual arguments to compute the result of the right hand side. Finally, we
return a tuple consists of a null value, a updated local environment with the updated binding of the left hand side $x$ as well as the updated memory store. 
The remaining cases are trivial.

$\ssaphi{\cdot}$ walks through the list of $\phi$ assignments. For
each $\phi$ of shape $x = \PHI(l_1:x_1,...,l_i:x_i,...,l_n:x_n)$, it
searches for the label matching with the incoming label $l_i$. The
value of $x_i$ will be assigned to the variable $x$.

The definitions of $\ssaassigns{\cdot}$, $\ssaassign{\cdot}$ and $\ssaexp{\cdot}$ are straight-forward and we omit the details.

\section{SSAFJ-EH to \fjlam Translation} \label{sec:translation}

\subsection{Syntax of \fjlam} \label{sec:target_syntax}
We consider the valid syntax of our target language FJ$_{\lambda}$
{\footnotesize
\bda{rccl}
\tlabel{CLASSDECL} & \CD & ::= & class~C~\{\overline{\FD}; \overline{\MD}  \}   \\
\tlabel{FIELDDECL} & \FD & ::= & T~F  \\
\tlabel{METHODDECL} & \MD & :: = & T~M~(T~X) \{ \overline{\VD};\overline{S} \}
\\
\tlabel{VARDECL} & \VD & :: = & T~X  \mid K~X = \lambda \\
\tlabel{STATEMENT} & S & ::= & A \mid \RET~E \mid \IF~ (E)~ \{ \overline{S} \}~ \ELSE~ \{ \overline{S} \} \\
\tlabel{ASSIGNMENT} & A & ::= & X = E \mid E.F = E \\
\tlabel{EXPRESSION} & E  & ::= & V \mid X \mid \THIS \mid X(\overline{E}) \mid E.M(E) \mid 
E.F
\\ & & & \mid E ~op~E \mid \NEW~T() \\
\tlabel{BASIC TYPE} & T & ::= & int \mid bool \mid void \mid
C \\
\tlabel{FUNCTION TYPE} & K & ::= & T \mid  K \Rightarrow K \\
\tlabel{VALUE} & V & ::= & c \mid \lambda \mid LOC \mid \NULL \\
\tlabel{MEMLOC} & LOC & ::= & \loc{0} \mid \loc{1} \mid ...  \\
\tlabel{LAMBDA} & \lambda & ::= & \overline{(K~X)} \arrow \{ \overline{S} \} 
\eda
}%

\fjlam is an extension of FJ with the support of
anonymous functions, i.e. lambda abstraction. \fjlam
differs from SSAFJ-EH as follows.
Labels, while loop, try-catch and throw
statements are excluded. 
\fjlam supports a limited form of higher order functions.
$K~X=\lambda$ defines a
local constant variable whose value is initialized to a lambda
abstraction $\lambda$.  
Lambda abstraction $\overline{(K~X)} \rightarrow \{ \overline{S}\}$ denotes an anonymous
function that expects zero or more parameters. Lambda functions do not
introduce local variables within their own scopes. Lambda functions
are nested in a top-level
method. \footnote{The examples given in
Section~\ref{sec:example} Figures~\ref{f:get_cps_flatten} and
\ref{f:combinators_flatten} seem to be violating this
restriction. The violation is due to the ``flattening'' effect, which
can be undone.} 
Note that $X$ and $Y$  denote variables. Variables 
declared in a method are accessible within its nested functions. 
$X(\overline{E})$ denotes a
function application where $X$ is a variable bound to a lambda
abstraction. $E.M(E)$ denotes a method application. 
The value $V$ in the target language
includes constants, lambda abstraction and memory locations.

\subsection{Semantics of \fjlam } \label{sec:target_semantics}
\figtwocol{f:target_semantic_p1}{Denotational Semantics of \fjlam} {
{\footnotesize
  \bda{rrcl}
  \tlabel{Global Decl Env} & {\tt GENV} & \subseteq & ({\tt CLASS} \times \overline{\tt FIELDDECL}) ~ \cup \\
  & & &  (({\tt CLASS} \times {\tt METHODNAME}) \times {\tt METHODDECL}) \\
  \tlabel{Local Decl Env} & {\tt LENV} & \subseteq & ({\tt VARIABLE} \times {\tt VALUE} ) \\
  \tlabel{Memory Store} & {\tt STORE} & \subseteq & ({\tt MEMLOC}
  \times {\tt OBJECT}) \\
  \tlabel{OBJECT} & \OBJ & ::= & \OBJ(T,\rho)  
  \eda
  
  \bda{ll}
 \multicolumn{2}{l}{\tarmeth{\cdot} :: {\tt METHODDECL} \arrow {\tt
     VALUE} \arrow {\tt VALUE} \arrow {\tt GENV} \arrow {\tt STORE}} \\
 \multicolumn{1}{l}{\bgap \bgap \bgap \arrow ({\tt VALUE},{\tt STORE})} \\ 
 \tarmeth{(T'~ M(T~X)\{\overline{\VD}; \overline{S}\})}\ V_0\ V_x\ genv\ st = \\
 \bgap \LET~lenv= \tarvdecl{\overline{\VD}}~\{(this,V_0), (X,V_x)\} \\
 \bgap \IN~ \CASE~ \tarstmt{\overline{S}}~genv~lenv~st~\OF ~ (V,\_ , st') \arrow (V, st') 
  \eda

\bda{l}
  \tarvdecl{\cdot} :: [{\tt VARDECL}] \arrow {\tt LENV} \arrow {\tt LENV} \\
  \tarvdecl{K~X=\lambda; \overline{\VD}}~lenv = \tarvdecl{\overline{\VD}}~(lenv+(X,\lambda))
  \eda

\bda{ll}
 \multicolumn{2}{l}{\tarstmt{\cdot} :: {\tt STATEMENT} \arrow {\tt GENV} \arrow {\tt LENV} \arrow {\tt STORE}
   \arrow ({\tt VALUE}, {\tt STORE}) } \\
 \tarstmt{[]}~genv~lenv~st = (\NULL, lenv, st) \\
 \tarstmt{A; \overline{S}}~genv~lenv~st = \LET~(V, lenv', st') = \tarassign{A}~genv~lenv~st \\
 \bgap \IN ~ \tarstmt{\overline{S}}~genv~lenv'~st' \\ 

 \tarstmt{\RET~E}~genv~lenv~st = \CASE~\tarexp{E}~genv~lenv~st~\OF \\
 \bgap (V, st') \arrow (V, lenv, st') \\

 \tarstmt{\IF~(E)~\{\overline{S_1}\} \ELSE \{\overline{S_2}\}}~genv~lenv~st = \CASE~\tarexp{E}~genv~lenv~st~\OF \\
 \bgap (true, st') \arrow \tarstmt{\overline{S_1}}~genv~lenv~st'\\
 \bgap (false, st') \arrow \tarstmt{\overline{S_2}}~genv~lenv~st' 
 \eda

 \bda{l}
 \tarassign{\cdot} :: {\tt ASSIGNMENT} \arrow {\tt GENV} \arrow {\tt LENV} \arrow {\tt STORE} \arrow ({\tt VALUE}, {\tt LENV}, {\tt STORE})
 \eda

 \bda{c}
  \tarexp{\cdot} :: {\tt EXPRESSION} \arrow {\tt GENV} \arrow {\tt
    LENV} \arrow {\tt STORE} \arrow ({\tt VALUE}, {\tt STORE})
  \eda
  \bda{ll}
  \tarexp{X(\overline{E})}\ genv\ lenv\ st = \CASE~lenv(X)~\OF \\
  \bgap ((\overline{K~X}) \rightarrow \{ \overline{S} \}) \arrow \CASE~\tarstmt{\overline{S}}\ genv\ lenv'\ st_n\ \OF \\
  \bgap \bgap (V, _, st') \arrow (V, st') \\
  \bgap \mbox{where}~(V_1,st_1) = \tarexp{E_1}\ genv\ lenv\ st \\
  \bgap \bgap ~~~... \\ 
  \bgap \bgap ~~~(V_n,st_n) = \tarexp{E_n}\ genv\ lenv\ st_{n-1} \\
  \bgap \bgap ~~~lenv' = lenv + (X_1, V_1) + ... + (X_n, V_n) \\ 
  \tarexp{E_1.M(E_2)}\ genv\ lenv\ st = \CASE~\tarexp{E_1}\ genv\ lenv\ st\ \OF \\
  \bgap (\loc{n}, st_1) \arrow \CASE~st_1(\loc{n})~\OF\\
  \bgap \bgap \OBJ(T,\rho) \arrow \CASE\ genv(T,M)\ \OF\\
  \bgap \bgap \bgap \MD \arrow \CASE\ \tarexp{E_n}\ genv\ lenv\ st_1\ \OF \\
  \bgap \bgap \bgap \bgap (V_2, st_2) \arrow \tarmeth{\MD}\ \loc{n}\ v_2\ genv\ st_2
  \eda
}
}
In Figure~\ref{f:target_semantic_p1} 
we describe the denotation semantics of \fjlam.
We use the upper case symbols {\tt GENV}, {\tt LENV} and {\tt STORE} to
capture the run-time bindings. They are similar to the counter-parts
found in the SSAFJ-EH.


$\tarmeth{\cdot}$ is similar to $\ssameth{\cdot}$ except that it
does not keep track of labels and exceptions. 
$\tarvdecl{\cdot}$ is nearly identical to $\ssavdecl{\cdot}$ except that it
handles an extra case of local function declaration. 

$\tarstmt{\cdot}$ is a simplified version of $\ssablks{\cdot}$ without the need of keeping track of
the labels and the exceptions. 

$\tarassign{\cdot}$ is nearly identical to $\ssaassign{\cdot}$, hence its definitions are omitted.

$\tarexp{\cdot}$ differs from $\ssaexp{\cdot}$ in case of function/method
application. There are two different scenarios.
\tlabel{I} In case of  $\tarexp{X(\overline{E})}$ where $lenv(X)$ yields a lambda abstraction.
We evaluate all the actual parameters $E_1$ to $E_n$ into $V_1$ to $V_n$ with the memory store being updated
and propagated. We create an extended local
environment by binding $X_i$s to $V_i$s. Finally we proceed with the evaluation the body of the  lambda abstraction
under the new environment and memory store.
\tlabel{II} In case of function application $\tarexp{E_1.M(E_2)}$, we evaluate $E_1$ into a memory location
$\loc{n}$ with an updated memory store $st'$. By looking up $st'(\loc{n})$ we retrieve the definition of
the method associated with name $M$. We then evaluate $E_2$ and apply the resulting value to the method. 

\subsection{SSAFJ-EH to \fjlam Translation using CPS} \label{sec:ssa2cpstrans}

\figtwocol{fig:ssa2cpstrans-p1}{SSAFJ-EH to \fjlam Translation (Part 1)} {

{\footnotesize
\bda{l}
\cpsconvmeth{\cdot} :: {\tt MethodDecl} \arrow {\tt METHODDECL} \\ 
\cpsconvmeth{t'~m~(t~x)\{\overline{vd}; \overline{b}\}} = \\
\bgap \LET~\overline{\VD}~=~\cpsconvvdecl{\overline{vd}} \\
\bgap ~~~~(\overline{\VD'},E)~=~\cpsconvblk{[x/input]\overline{b}}~[]~[] \\
\bgap ~~~~ T = t; T' = t'; X = x ; M = m \\ 
\bgap ~~~~ D = T \Rightarrow (Exception \Rightarrow void) \Rightarrow (T' \Rightarrow void) \Rightarrow void ~ M_{cps} =  \\ 
\bgap \bgap  (T~in) \arrow (Exception \Rightarrow void~raise) \arrow (T' \Rightarrow void~k) \arrow \\
\bgap \bgap \bgap \{ input = in; E(raise)(()\arrow k(res)) \};\\
\bgap \IN ~ T'~M~(T~X)~\{ \overline{\VD} ++ \overline{VD'} ++ [D]; T~input; T'~res; Exception~ex; \\
\bgap \bgap M_{cps}(X)(id_{raise})(r \arrow \{res = r; return;\}); return~res; \} 
\eda

\bda{l}
\cpsconvvdecl{\cdot} :: {\tt [VarDecl]} \arrow {\tt [VARDECL]} \bgap \bgap \bgap \bgap \bgap \bgap
\eda

\bda{l}
\cpsconvblk{\cdot} :: {\tt [Block]} \arrow {\tt [Phi]} \arrow {\tt [Phi]} \arrow ({\tt [VARDECL]}, {\tt EXPRESSION}) \\

\cpsconvblk{l:\{ \IF~(e)~\{\overline{b'}\} \ELSE \{\overline{b''}\}~\JOIN~\{\overline{\phi}\}\}} ~\overline{\phi_k}~ \overline{\phi_r} = \\
\bgap \LET~(\overline{D'},E')=~\cpsconvblk{\overline{b'}}~ \overline{\phi}~ \overline{\phi_r}  \\
\bgap ~~~~ (\overline{D''},E'')=~\cpsconvblk{\overline{b''}}~ \overline{\phi}~ \overline{\phi_r} \\
\bgap ~~~~ E =~\cpsconvexp{e} \\
\bgap ~~~~ (\overline{D'''},E''')=~\cpsconvjump{\overline{\phi_k}}~l \\
\bgap \IN (\overline{D'} ++ \overline{D''} ++ \overline{D'''}, seq(ifelse(() \arrow E, E', E''), E''')) \\

\cpsconvblk{l:\{ \IF~(e)~\{\overline{b'}\} \ELSE \{\overline{b''}\}~\JOIN~\{\overline{\phi}\}\}; \overline{b} } ~\overline{\phi_k}~ \overline{\phi_r} = \\
\bgap \LET~(\overline{D'},E')=~\cpsconvblk{\overline{b'}}~ \overline{\phi}~ \overline{\phi_r} \\
\bgap ~~~~ (\overline{D''},E'')=~\cpsconvblk{\overline{b''}}~ \overline{\phi}~ \overline{\phi_r} \\
\bgap ~~~~ E =~\cpsconvexp{e} \\
\bgap ~~~~ (\overline{D'''},E''')=~\cpsconvblk{\overline{b}}~\overline{\phi_k}~\overline{\phi_r} \\
\bgap \IN (\overline{D'} ++ \overline{D''} ++ \overline{D'''}, seq(ifelse(() \arrow E, E', E''), E''')) \\

\cpsconvblk{ l: \{ \JOIN~\{\overline{\phi}\}~\WHILE~(e)~\{\overline{b'}\} \}}~ \overline{\phi_k}~ \overline{\phi_r} = \\
\bgap \LET~(\overline{D}, E)=\cpsconvjump{\overline{\phi}}~minLabel(\overline{\phi}) \\
\bgap ~~~~ E' = \cpsconvexp{e} \\
\bgap ~~~~ (\overline{D''},E'')=~\cpsconvblk{\overline{b'}}~\overline{\phi}~\overline{\phi_r} \\
\bgap ~~~~ (\overline{D'''}, E''')=~\cpsconvjump{\overline{\phi_k}}~l \\
\bgap \IN~(\overline{D} ++ \overline{D''} ++ \overline{D'''}, seq(E,loop( () \arrow E', E'', E'''))) \\

\cpsconvblk{ l: \{ \JOIN~\{\overline{\phi}\}~\WHILE~(e)~\{\overline{b'}\} \} ; \overline{b}} ~\overline{\phi_k}~\overline{\phi_r} = \\
\bgap \LET~(\overline{D}, E)=\cpsconvjump{\overline{\phi}}~minLabel(\overline{\phi}) \\
\bgap ~~~~ E' = \cpsconvexp{e} \\
\bgap ~~~~~(\overline{D''},E'')=~\cpsconvblk{\overline{b'}}~\overline{\phi}~\overline{\phi_r} \\
\bgap ~~~~ (\overline{D'''}, E''')=~\cpsconvblk{\overline{b}}~\overline{\phi_k}~\overline{\phi_r} \\
\bgap \IN~(\overline{D} ++ \overline{D''} ++ \overline{D'''}, seq(E,loop( () \arrow E', E'', E'''))) \\

\cpsconvblk{ l: \{ \THROW~e \} }~\overline{\phi_k}~ \overline{\phi_r}
= \LET~\overline{(X,E)} = \cpsconvphi{\overline{\phi_r}}~l \\
\bgap ~~~ D = (Exception \Rightarrow void) \Rightarrow (void
\Rightarrow void) \Rightarrow void~m_l = \\
\bgap \bgap ~~~~ (Exception \Rightarrow
void~raise) \arrow (void \Rightarrow void ~k) \arrow  \\
\bgap \bgap ~~~~ \{\overline{X = E}; \RET~raise(\cpsconvexp{e}); \} \\
\bgap \IN~ ([D], m_l) \\

\cpsconvblk{ l: \{ \RET~e\} } \overline{\phi_k}~\overline{\phi_r} =
\LET~E = \cpsconvexp{e} \\
\bgap~~~ D = (Exception \Rightarrow void) \Rightarrow (void
\Rightarrow void) \Rightarrow void m_l = \\
\bgap \bgap ~~~~ (Exception \Rightarrow void~raise) \arrow (void
\Rightarrow void~k) \arrow \\
\bgap \bgap ~~~~ \{ res = E; \RET~k();\} \\
\bgap \IN~([D], m_l) \\

\cpsconvblk{ l: \{ \TRY \{ \overline{b} \} \JOIN \{ \overline{\phi'_r} \}
~\CATCH~(t~x) \{ \overline{b'} \} \JOIN~\{\phi'_k\} \} }
~\overline{\phi_k}~\overline{\phi_r} = \\
\bgap \LET~(\overline{D}, E) =
\cpsconvblk{\overline{b}}~\overline{\phi'_k}~\overline{\phi_r'} \\
\bgap ~~~~ (\overline{D'},E') =
\cpsconvblk{\overline{b'}}~\overline{\phi'_k}~\overline{\phi_r} \\
\bgap ~~~~ E'' = (Exception~x) \arrow \{ ex = x; \RET~E'; \} \\
\bgap ~~~~ (\overline{D'''},E''') =
\cpsconvjump{\overline{\phi_k}}~l \\
\bgap \IN ~(\overline{D} ++ \overline{D'} ++ \overline{D'''},
seq(trycatch(E,E''), E''')) \\

\cpsconvblk{ l: \{ \TRY \{ \overline{b} \} \JOIN \{ \overline{\phi'_r} \}
~\CATCH~(t~x) \{ \overline{b'} \} \JOIN~\{\phi'_k\} \} ; \overline{b''} }
~\overline{\phi_k}~\overline{\phi_r} = \\
\bgap \LET~(\overline{D}, E) =
\cpsconvblk{\overline{b}}~\overline{\phi'_k}~\overline{\phi_r'} \\
\bgap ~~~~ (\overline{D'},E') =
\cpsconvblk{\overline{b'}}~\overline{\phi'_k}~\overline{\phi_r} \\
\bgap ~~~~ E'' = (Exception~x) \arrow \{ ex = x; \RET~E'; \} \\
\bgap ~~~~ (\overline{D'''},E''') =
\cpsconvblk{\overline{b''}}~\overline{\phi_k}~\overline{\phi_r} \\
\bgap \IN ~(\overline{D} ++ \overline{D'} ++ \overline{D'''},
seq(trycatch(E,E''), E''')) 

\eda
}
}

\figtwocol{fig:ssa2cpstrans-p2}{SSAFJ-EH to \fjlam Translation (Part 2)} {
  {\footnotesize
    \bda{l}
\cpsconvblk{ l: \{ \overline{a} \}}~\overline{\phi_k}~\overline{\phi_r} = \LET~\overline{A} =
\cpsconvassign{\overline{a}} \\
\bgap ~~~~ D = (Exception \Rightarrow void) \Rightarrow (void
\Rightarrow void) \Rightarrow void~m_l = \\
\bgap \bgap  ~~~~ (Exception \Rightarrow
void~raise) \arrow (void \Rightarrow void ~k) \arrow \\
\bgap \bgap  ~~~~ \{ \overline{A}; \RET~k(); \} \\
\bgap ~~~~ (\overline{D'}, E') = \cpsconvjump{\overline{\phi_k}}~l \\ 
\bgap \IN ([D] ++ \overline{D'}, seq(m_l, E') \\

\cpsconvblk{l:\{\overline{a} \} ; \overline{b} }~\overline{\phi_k}~\overline{\phi_r}
= \LET~\overline{A} = \cpsconvassign{\overline{a}} \\
\bgap ~~~~ D = (Exception \Rightarrow void) \Rightarrow (void
\Rightarrow void) \Rightarrow void~m_l = \\
\bgap \bgap  ~~~~ (Exception \Rightarrow
void~raise) \arrow (void \Rightarrow void ~k) \arrow \\
\bgap \bgap  ~~~~ \{ \overline{A}; \RET~k(); \} \\
\bgap ~~~~(\overline{D'},E') =
\cpsconvblk{\overline{b}}~\overline{\phi_k}~\overline{\phi_r}  \\
\bgap \IN~  ([D] ++ \overline{D'}, seq(m_l, E')) \\

\cpsconvblk{l: \{ x = e_1.m(e_2) \}}~\overline{\phi_k}~\overline{\phi_r}
  = \LET~E_1 = \cpsconvexp{e_1} \\
  \bgap ~~~~ E_2 = \cpsconvexp{e_2} \\
  \bgap ~~~~ D = (Exception \Rightarrow void) \Rightarrow (void
  \Rightarrow void) \Rightarrow m_l = \\
  \bgap \bgap ~~~~(Exception \Rightarrow void ~raise) \arrow (void
  \Rightarrow void~k) \arrow \\
  \bgap \bgap ~~~~ \{ E_1.m_{cps}(E_2)(raise)( (T~v) \arrow \{ x = v;
  \RET~k(); \}) \} \}\\
  \bgap ~~~~ (\overline{D'}, E') = \cpsconvjump{\overline{\phi_k}}~l\\
  \bgap \IN~([D] ++ \overline{D'}, seq(m_l,E')) \\

\cpsconvblk{l: \{ x = e_1.m(e_2);\overline{b} \}}~\overline{\phi_k}~\overline{\phi_r}
  = \LET~E_1 = \cpsconvexp{e_1} \\
  \bgap ~~~~ E_2 = \cpsconvexp{e_2} \\
  \bgap ~~~~ D = (Exception \Rightarrow void) \Rightarrow (void
  \Rightarrow void) \Rightarrow m_l = \\
  \bgap \bgap ~~~~(Exception \Rightarrow void ~raise) \arrow (void
  \Rightarrow void~k) \arrow \\
  \bgap \bgap ~~~~ \{ E_1.m_{cps}(E_2)(raise)( (T~v) \arrow \{ x = v;
  \RET~k(); \}) \} \}\\
  \bgap ~~~~ (\overline{D'}, E') =
  \cpsconvblk{\overline{b}}~\overline{\phi_k}~\overline{\phi_r} \\
  \bgap \IN~([D] ++ \overline{D'}, seq(m_l,E')) 
  \eda

  \bda{l}
\cpsconvjump{\cdot} :: [{\tt Phi}] \arrow Label \arrow ([{\tt
  VARDECL}], {\tt EXPRESSION}) \\
\cpsconvjump{\overline{\phi}}~l = \LET~\overline{(X,E)} =
\cpsconvphi{\overline{\phi}}~l \\
\bgap~~~~ D = (Exception \Rightarrow void) \Rightarrow (void
\Rightarrow void) \Rightarrow void~ mk_l = \\
\bgap \bgap ~~~~ (Exception \Rightarrow void~raise) \arrow (void
\Rightarrow void~k) \arrow \\
\bgap \bgap ~~~~ \{ \overline{X = E}; \RET~k(); \} \\
\bgap \IN ~ ([D], mk_l)
\eda

\bda{l}
\cpsconvassign{\cdot} :: [{\tt Assignment}] \arrow [{\tt ASSIGNMENT}] \\
\eda

\bda{l}
\cpsconvexp{\cdot}:: {\tt Expression} \arrow {\tt EXPRESSION} \\
\eda

\bda{l}
\cpsconvphi{\cdot}:: [{\tt Phi}] \arrow Label \arrow [({\tt VARIABLE},
{\tt EXPRESSION})] \\
\cpsconvphi{[]}~\_~ = [] \\
\cpsconvphi{x = phi(..., l_i:x_i, ...); \overline{\phi}}~ l  \mid l ==
l_i = (x, x_i): \cpsconvphi{\overline{\phi}}~l
\eda
}
}
We describe the SSAFJ-EH to \fjlam translation using CPS in 
Figures~\ref{fig:ssa2cpstrans-p1} and~\ref{fig:ssa2cpstrans-p2}.
Specifically, we use command-based continuation pass style.

There are mainly two types of
continuations, the exception continuation {\tt Exception => void} and the normal continuation {\tt void => void}.
Each function in CPS form expects the first argument as the exception continuation and the second one as the
normal continuation, except for the top level method. 

The function $\cpsconvmeth{\cdot}$ converts a method from SSAFJ-EH
to \fjlam using CPS. 
Given an input method in SSAFJ-EH has type {\tt t => t'},
the conversion synthesizes the output (or translated) method of type
{\tt T => (Exception => void) => (T' => void) => void}, by letting $T = t$
and $T' = t'$. The first argument is the input, the second argument is
an exception continuation,
and the third argument is the normal continuation. The conversion
consists of the following steps.
Firstly we apply the helper function $\cpsconvvdecl{\cdot}$ to translate the
local variable declarations. $\cpsconvvdecl{\cdot}$ is an identity function,  we omit its details.
As the second step, we apply the helper function $\cpsconvblk{\cdot}$ which translates the list of blocks
from the source method. The result of the translation is a pair
consisting of a list of local lambda declarations and a main
expression. The main expression $E$ is then applied to the exception
continuation $raise$ and the normal continuation $()\arrow k(res)$.
At last we synthesize the public interfacing  method $M$ which wraps around the CPS counter-part
$M_{cps}$. \footnote{The continuation $r \arrow \{res = r; return;\}$ could have been simplified
  to $r \arrow \{ return; \}$. However we keep to former just for consistency.}

Most of the translation tasks are computed in the helper function
$\cpsconvblk{\cdot}$. The function expects a list of blocks, a list of
$\phi$ assignments from the subsequent block in the normal continuation and a list of
$\phi$ assignments from the subsequent block in the exception continuation.
$\cpsconvblk{\cdot}$ translates the blocks structurally. 

\begin{itemize}
\item In case of a singleton list containing an if-else block, we
  apply a helper function $\cpsconvexp{\cdot}$ to translate the
  conditional expression. Then we apply $\cpsconvblk{\cdot}$
   recursively to the blocks from the then-branch and the
   else-branch by using the $\phi$ assignments, $\overline{\phi}$,
   from the if-else statement's join clause.  To ``connect'' the
   translated if-else back to the subsequent block in the normal continuation,
   we apply another helper function
   $\cpsconvjump{\cdot}$ to construct a continuation that resolves
   $\overline{\phi_k}$ with respect to $l$.
   The main expression is constructed structurally from
   the derived expressions from the various sub-steps with $seq$ and
   $ifelse$ combinators, whose definitions can be found in Figure~\ref{f:combinators_flatten}.
\item In case of a non singleton list of which the head is an if-else block,
  we perform a trick similar to the previous case, except that we do
  not construct a continuation with $\cpsconvjump{\cdot}$ to resolve
  $\overline{\phi}_k$. Instead we
  apply $\cpsconvblk{\cdot}$ to $\overline{b}$ recursively.
\item In case of a singleton list containing a while block, we first need to
  apply $\cpsconvjump{\cdot}$ to resolve the $\overline{\phi}$ with respect to the
  label of the block from which we enter the while loop.
  For convenience, we assume that there exists
  a partial order among labels, i.e. $L_i \prec L_j$ implies that $L_i$ must be on the
  path leading from $L_o$ to $L_j$, where $L_0$ is the method's entry label. We assume that
  there are only two labels in the $\overline{\phi}$ assignments in all while blocks, i.e.
  the first label is the entry label to the while block, and the
  second label is loop-back label, and $minLabel(\overline{\phi})$
  returns the entry label. 
  Such a restrictive form does not limit the expressiveness of the language.
  We assume that there exists a pre-processing step that convert any programs into this form.

  After resolving $\overline{\phi}$ with respect to the 
  entry label to the while block, we apply $\cpsconvblk{\cdot}$ to $\overline{b}$ recursively to translate the while body.
  Lastly we apply $\cpsconvjump{\cdot}$ to construct a continuation that resolves
   $\overline{\phi}_k$ with respect to $l$. We build the main expression using
   the $seq$ and $loop$ combinators,
   whose definitions can be found in Figure~\ref{f:combinators_flatten}.
\item In case of a singleton list containing a try-catch block, we
  apply $\cpsconvblk{\cdot}$ recursively to the block in the try
  clause $\overline{b}$ with $\overline{\phi_k'}$ as $\phi$
  assignments from the normal continuation
  and $\overline{\phi_r'}$ from the exception continuation.
  The catch clause block $\overline{b'}$ is translated with
  $\overline{\phi_k'}$ as $\phi$ assignments from the normal continuation and
  $\overline{\phi_r'}$ from the exception continuation. In order to bind the
  exception into the variable $x$, we define a wrapper lambda
  expression which expects the exception as input and assigns it to
  $ex$. (Recall that $ex$ is defined in the top level method). Lastly,
  we construct a connecting continuation by resolving
  $\overline{\phi_k}$ with the current label $l$.
\item In case of a singleton list containing a throw block, we
  first resolve the $\overline{\phi_r}$ from  the exception handler
  with respect to the current label $l$
  by calling $\cpsconvphi{\cdot}$. Taking the result from the $\phi$
  resolution, we define a continuation
  function $m_l$ in which we bind the results, and call the exception
  continuation $raise()$ with the translation of the $e$.
\item In case of a singleton list containing a return block,  we
  define a continuation function $m_l$ in which we assign the
  translation of $e$ to $res$. (Recall that $res$ is defined in the
  top level method). Then we call the continuation $k$.

\item In cases of a list with a method invocation as the head, we
  translate the sub-expressions $e_1$ and $e_2$ into $E_1$ and $E_2$.
  We define a continuation function $m_l$ in which we invoke
  $E_1.m_{cps}(E_2)$ with $raise$ as the exception continuation and the
  normal continuation is a lambda expression that captures the result
  of the method invocation into an argument $v$. In the body of the
  lambda exression we assign $v$ to $x$ before invoking the continuation $k$.
  Note that we treat $M_{cps}$ same as $m_{cps}$ and the call of $m_{cps}$ could a recursive call or
  another method sharing the same closure context in the same scope.
\end{itemize}
The rest of the $\cpsconvblk{\cdot}$ cases are trivial. 

The helper function $\cpsconvphi{\cdot}$ takes a list of $\phi$ assignments, a
label and returns a list pair variable-expression pairs. For each $\phi$
assignment, it picks the right $x_i$ associated with the matching
label $l$ as the second component of the resulting pair.

The helper function $\cpsconvjump{\cdot}$ synthesizes a continuation
function that connects the block with label $l$ with the block that
$\overline{\phi}$ is defined, by making use of  $\cpsconvphi{\cdot}$.

Helper functions $\cpsconvassign{\cdot}$ and $\cpsconvexp{\cdot}$ are
identity functions, whose definitions are omitted.

\lstset
{ 
    language=java,
    basicstyle=\footnotesize,
    numbers=none,
    stepnumber=1,
    showstringspaces=false,
    tabsize=1,
    breaklines=true,
    breakatwhitespace=false,
}

\begin{figure}
  \begin{lstlisting}
int get(int x) {
  int i_1, i_2, i_5, i_6, t_6, r_1, r_2. r_7;
  int input, res; Exception ex;
    
  int => ExCont => (int => void) => void get_cps =
  x -> raise -> k -> {
    input = x;
    return seq(get1, seq(trycatch
      (seq(ifelse(n-> {input < i_1},
           get4,seq(getk3b, loop( n->{i_5<input}, seq(get6,get6k), seq(get7,get7k)))), getk3a)
      , e -> {ex = e; return seq(get8, get8k);}), get9)
      )(raise)(n->k(res))
    }
  ExCont => NmCont => void get1 = 
      (ExCont raise) -> (NmCont k) -> {
          i_1 = this.lpos; r_1 = -1; return k();
  }
  ExCont => NmCont => void getk3a = raise -> k
    -> {r_2 = r_7; return k();}
  ExCont => NmCont => void get4 = raise -> k 
    -> {i_2 = i_1; raise(new Exception()})
  ExCont => NmCont => void getk3b = raise -> k
    -> {i_5 = i_1; return k();}
  ExCont => NmCont => void get6 = raise -> k
    -> { t_6 = this.f1 + this.f2; this.f1 = this.f2;
         this.f2 = t_6; i_6 = i_5 + 1; return k();} 
  ExCont => NmCont => void getk6 = raise -> k
    -> {i_5 = i_6; return k();}
  ExCont => NmCont => void get7 = raise -> k 
    -> { this.lpos = i_5; r_7 = this.f2; return k();}
  ExCont => NmCont => void get7k = rise -> k
    -> { i_2 = i_5; return k() }
  ExCont => NmCont => void get8 = raise -> k 
    -> { System.out.println("..."); return k();}
  ExCont => NmCont => void get8k = raise -> k 
    -> { r_2 = r_1; return k();}
  ExCont => NmCont => void get9 = raise -> k
    -> { res = r_2; return k(); }
  get_cps(x)(id_raise)(i -> res = i; return);
  return res;
}

\end{lstlisting}
\hrule
\vspace{-3mm}
\caption{SSA to CPS Translation of {\tt fib}} \label{f:cps-translated-fib}
\end{figure}
In Figure~\ref{f:cps-translated-fib}, we find the full CPS translation
of the {\tt get} method in Figure~\ref{f:get_ssa}.
The result should be identical to the one in
Figure~\ref{f:get_cps_flatten}, except that we do not apply
``flattening'' to nested and curry function calls, we insert extra
connection blocks thanks to the $\phi$ resolutions.  

\begin{definition} [Consistent Global Environments] \label{def:consistent-genv}
Let $genv \in {\tt GEnv}$ and $genv' \in {\tt GENV}$. Then we say
$genv \turns genv'$ iff $\forall (C,m) \in dom(genv) : genv'(C,m) = \cpsconvmeth{genv(C,m)}$.
\end{definition}
\begin{lemma} [SSAFJ-EH to \fjlam Translation Consistency] \label{lem:conistent-cps-trans}
  Let $m$ be a SSAFJ-EH method of a class $C$, $o$ be (a reference to) an object of
  class $C$, $v$ be a value such that $o.m(v)$ is
  well-typed and terminating. Let $M = \cpsconvmeth{m}$. Let $genv \in {\tt GEnv}$,
  $genv' \in {\tt GENV}$ such that $genv \turns genv'$. Then we have
  $\ssameth{m}~o~v~genv~\{\} = \tarmeth{M}~o~v~genv'~\{\}$.
\end{lemma}

\section{Obfuscation Potency Analysis} \label{sec:analysis}

We analyze the potency of the CPS-based control flow obfuscation.
\ignore{
We put on the hat of an attacker who gains access to the source code from the byte-codes
by applying decompilation tools such as {\tt javap} \cite{oracle-java} and attempts to recover the
control flow of the original source code using static analysis techniques such as
the inter-procedural analysis. 
}

Let's try to apply inter procedural control flow analysis to the obfuscated code in Figures~\ref{f:get_cps_flatten}
and \ref{f:combinators_flatten}.
\ignore{
For ease of analysis, we implicitly rewrite return statement
{\tt return e;} into 

\begin{lstlisting}
  var r_e = e;
  return r_e;
\end{lstlisting}
where {\tt r\_e} is a fresh variable.
}
Recall that the goal of the control flow analysis is to approximate
the set of possible lambda abstractions that a program variable
may capture during the run-time.
From that result, as an attacker, we can create a global control flow graph
with all the lambdas and methods involved.

Let $\alllambda$ denote the set of all possible lambda values in the obfuscated program in \fjlam. We have the
following lattice $(2^{\alllambda}, \subseteq)$, whose top element $\top$
is $\alllambda$ and $\bot$ is the empty set. We define the abstract
state of the analysis as a map lattice mapping variables to sets of lambda
functions. 

\bda{rccl}
\tlabel{STATE} & \sigma  & \subseteq & ({\tt VARIABLE} \times 2^{\alllambda}) 
\eda

\subsection{Context Insensitive Control Flow Analysis} \label{sec:context-insensitive}

We define the flow function $\flowftwo{\cdot}{\cdot} :: {\tt STATEMENT}
\rightarrow {\tt STATE} \rightarrow {\tt STATE}$. The flow function
takes a statement and a state and returns an updated state. 
Given a statement $S$, we write $\flowf{S}$ to denote
$\flowftwo{S}{\sigma_S}$ by making $\sigma_S$ an implicit argument where
$$\sigma_S = join(S) $$

Let $S$ be a statement and $pred(S)$ denote the set of preceding
statements of $S$, we define the join function $join(S)$ as 
$$ join(S) = \bigcup_{P \in pred(S)} \flowf{P}$$

The definition of flow function is given as follows.

\bda{rcl}
\flowftwo{ \RET~ E } {\sigma_S}  & =  & \sigma_S \\
\flowftwo{ \IF ~E~\{ S \}~ \ELSE ~ \{ S' \} }{\sigma_S} & = & \sigma_S  \\
\flowftwo{ E_1.F = E_2}{\sigma_S} & = & \sigma_S  \\
\flowftwo{ X = c }{\sigma_S} & = & \sigma_S - X \cup [ X \mapsto
  \emptyset ]  \\
\flowftwo{ X = E.F }{\sigma_S} & = & \sigma_S - X \cup [ X \mapsto
\emptyset ]  \\
\flowftwo{ X = E ~op ~E'}{\sigma_S} & = &   \sigma_S - X \cup [ X
  \mapsto \emptyset ]
\\
\flowftwo{ X = \NEW~T() } { \sigma_S } & = & \sigma_s - X \cup [ X \mapsto \emptyset ]
\\
\flowftwo{ X = \lambda } {\sigma_S}  & = & \sigma_S - X \cup [ X \mapsto 
\{ \lambda \} ] \\
\flowftwo{ X = Y } {\sigma_S} & = & \sigma_S - X \cup [ X \mapsto
\sigma_S(Y) ] \\
\flowftwo{ X = G(E_1, ..., E_n) } { \sigma_S } & = & \sigma_S - X \cup
[ X \mapsto {\tt returned} ]
\\
\mbox{where} \\
{\tt returned} & = & \bigcup_{\lambda \in \sigma_S(G)} \flowf{
  \retstmt{\lambda} } \\

\flowftwo{ X = E_1.M(E_2) } { \sigma_S } & = & \sigma_s - X \cup [ X
\mapsto {\tt returned} ]
\\ 
\mbox{where} \\
T & = & typeof(E_1) \\ 
{\tt returned} & = & \bigcup_{\lambda \in {\tt GENV}(T,M)} \flowf{
  \retstmt{\lambda} } \\
\eda
The first three cases handle return statement, if statement and field update. They do
not contribute any changes to abstract state.  In the cases of constant
assignment, field assignment, binary operator and  object instantiation
we update the variable $X$ with an empty set. In the case of lambda assignment, we set $X$ to be a
singleton set. In case of variable aliasing assignment, we set $X$'s
mapping to the same as the rhs. 
In case of lambda function invocation, we update the mapping of the variable $X$ with a union
of all returnable states from all the possible bindings of the
variable $G$ which is bound to some lambda expressions. In case of
method call, it is similar to the lambda function except that we look
up the lambda expression from the global environment.

We overload the flow function for a lambda function declaration, whose
output abstract state, will serve as the predecessor of the first statement in the function
body. 
\ignore{
\bda{l}
\flowf{ \lambda }  =  \\
\bigcup_{S \in \caller{\lambda}} \bot [a_1 \mapsto
\eval{\flowf{S}}{E_1^S}, ..., a_n \mapsto \eval{\flowf{S}}{E_n^S} ] \\
\\
\mbox{where} ~\{a_1, ..., a_n\}  = \farg{\lambda} \\ 
\eda
}

$$
\flowf{ \lambda }  =  \bigcup_{S \in \caller{\lambda}} \bot [a_1 \mapsto
\eval{\flowf{S}}{E_1^S}, ..., a_n \mapsto \eval{\flowf{S}}{E_n^S} ] \\
$$
where $\{a_1, ..., a_n\}  = \farg{\lambda}$.
Given a statement $S$ that calls $\lambda$, $E_i^S$ , denotes the actual argument at $i$th position.

The helper function
$\caller{\cdot} :: \Lambda \rightarrow [{\tt STATEMENT}]$ returns the set statements in which the function
$\lambda$ is invoked. Let $\bar{\sigma}$ denotes all the abstract
states collected from all the statements of the target program.

\bda{rcl}
\caller{\lambda} & = & \{ S | S \in {\tt STATEMENT} \wedge X \in
dom(\bar{\sigma}) \wedge \\
& & \lambda \in \sigma_S(X)  \wedge X(\overline{E}) \in
rhs(S)~\mbox{for some}~ \overline{E} \}
\eda
Helper function $\eval{\cdot}{\cdot} :: {\tt STATE} \rightarrow {\tt
  EXPRESSION} \rightarrow 2^{\Lambda}$, takes an abstract state and
returns a set of lambda functions which the expression might evaluate
to.
\bda{rcl}
\eval{\sigma}{c} & = & \emptyset \\
\eval{\sigma}{\lambda} & = & \{ \lambda \} \\
\eval{\sigma}{X} & = & \sigma(X) \\
\eval{\sigma}{E~op~E'} & = & \emptyset \\
\eval{\sigma}{E.F} & = & \emptyset \\
\eval{\sigma}{\NEW~T()} & = & \emptyset \\ 
\eda

We apply the above analysis to our running example in
Figures~\ref{f:get_cps_flatten} and \ref{f:combinators_flatten} until
the abstract state reaches the fix point.
We observe the following results. \\
{\footnotesize
  \begin{tabular} {  | c  c | c c | c  c | }
    \hline
  var & func & var & func & var & func \\ \hline
  ${\tt raise}_7$ & $\lambda_{37}$ & ${\tt k}_7$ & $\lambda_{37}$ &  ${\tt get}_2$ & $\lambda_{78}$ \\
  ${\tt get3}$ & $\lambda_{90}$ & ${\tt get5}$ & $\lambda_{52}$ & ${\tt get\_1\_2}$ & $\lambda_{68}$ \\
  ${\tt pseq}$ & $\lambda_{68}$ &  {\tt pseq\_raise} & $\lambda_{68}'$ & ${\tt raise_{18}}$ & $\lambda_{43}$\\
  ${\tt k_{18}}$ & $\lambda_{70}$ & ${\tt hdl_{23}}$ & $\lambda_{23}$ & ${\tt raise_{26}}$ & $\lambda_{79}$ \\
  ${\tt k_{26}}$ & $\lambda_{70}$ & ${\tt raise_{28}}$ & $\lambda_{79}$ & ${\tt k_{28}}$ & $\lambda_{55}$ \\
  ${\tt raise_{31}}$ & $\lambda_{79}$ & ${\tt k_{31}}$ & $\lambda_{70}$ & ${\tt raise_{33}}$ & $\lambda_{43}$ \\
  ${\tt k_{33}}$ & $\lambda_{70}$ & ${\tt raise_{35}}$ & $\lambda_{43}$ & ${\tt k_{35}}$ & $\lambda_{15}$ \\
  ${\tt cond_{50}}$ & $\lambda_8$ & {\tt visitor} & $\lambda_{28}$ & {\tt exit} & $\lambda_{31}$ \\
  ${\tt raise_{52}}$ & $\lambda_{79}$ & ${\tt k_{52}}$ & $\lambda_{70}$ & {\tt visitor\_raise} & $\lambda_{28}'$ \\
  {\tt ploop} & $\lambda_{50}$ & {\tt ploop\_raise} & $\lambda_{52}'$ & {\tt exit\_raise} & $\lambda_{31}'$ \\
 ${\tt raise_{68}}$ & $\lambda_{43}$ &  {\tt first} & $\lambda_{18}, \lambda_{78}$ & {\tt second} & $\lambda_{35}, \lambda_{68}$ \\ 
  ${\tt k_{68}}$ & $ \lambda_{15}$ & {\tt first\_raise} & $\lambda_{18}', \lambda_{78}'$ & {\tt second\_raise} & $\lambda_{35}', \lambda_{68}'$ \\
  ${\tt raise_{78}}$ &  $\lambda_{43}$ & ${\tt k_{78}}$ & $\lambda_{70}$ & {\tt hdl\_ex} & $\lambda_{33}$  \\
  {\tt tr\_hdl} & $\lambda_{90}'$ & {\tt hdl\_ex\_raise} & $\lambda_{33}'$ & ${\tt cond_{88}}$ & $\lambda_9$ \\
  {\tt th} & $\lambda_{26}$ & {\tt el} & $\lambda_{52}$ & ${\tt raise_{90}}$ & $\lambda_{79}$ \\
  ${\tt k_{90}}$ & $\lambda_{70}$ & {\tt th\_raise} & $\lambda_{26}'$ & {\tt el\_raise} & $\lambda_{52}'$ \\
  {\tt id\_bind} & $\lambda_{37}$ & {\tt get\_x} & $\lambda_7'$ & {\tt get\_x\_raise} & $\lambda_7''$ \\
  {\tt tr} & $\lambda_{90}$ & ${\tt hdl_{77}}$ & $\lambda_{23}$ & {\tt get\_cps} & $\lambda_7$ \\
  ${\tt cond3}$ & $\lambda_9$ & ${\tt cond5}$ & $\lambda_8$ & ${\tt get1}$ & $\lambda_{18}$ \\
  ${\tt get4}$ & $\lambda_{26}$ & ${\tt get6}$ & $\lambda_{28}$ & ${\tt get7}$ & $\lambda_{31}$ \\
  ${\tt get8}$ & $\lambda_{33}$ & ${\tt get9}$ & $\lambda_{35}$ & {\tt id\_raise} & $\lambda_{43}$ \\
  {\tt n\_loop} & $\lambda_{55}$ & {\tt n\_second} & $\lambda_{70}$ & {\tt loop} & $\lambda_{50}$ \\
  {\tt seq} & $\lambda_{67}$ & {\tt trycatch} & $\lambda_{77}$ & {\tt ex\_hdl} & $\lambda_{79}$ \\
  {\tt ifelse} & $\lambda_{88}$ & {\tt n\_k\_res} & $\lambda_{15}$  & & \\ \hline
\end{tabular}
}
\\
For clarity and brevity, we adopt the following naming convention. We add line numbers to make common variables unique,
e.g. ${\tt raise_7}$ denotes the {\tt raise} from line 7.

As we can observe from the above, most of variables are given a unique lambda term to which they can be bound, except for
{\tt first}, {\tt second}, {\tt first\_raise} and {\tt second\_raise}. This is caused by the fact that the function {\tt seq}
is invoked in two different locations.

Through the analysis result, we can approximate a caller-callee relation between
lambda abstractions. We reconstruct a global CFG of the obfuscated {\tt get} by combining the
call graphs and the local control flow graphs. The resulting CFG is presented in Figure~\ref{f:cfg_obfs}.

As we discuss in the earlier section, the loss of precision is caused
by the incompleteness of the context sensitive control flow
analysis. 

\subsection{Context Sensitive Control Flow Analysis}

A smarter attacker may attempt to uncover the CFG with better precision
with context sensitive analysis.

In context sensitive analysis, we extend the abstract state with a context.
\bda{rccl}
\tlabel{STATE} & \sigma  & \subseteq & ({\tt Variable} \times 2^{\alllambda}) \cup \{\unrec\}
\eda
In this lattice, $\unrec$ is the new $\bot$. 

We redefine the flow function $\flowfthree{\cdot}{\cdot}{\cdot} :: {\tt STATEMENT} \rightarrow {\tt CONTEXT} \rightarrow {\tt STATE} \rightarrow {\tt STATE}$.
The flow function takes a statement, a context and a state and returns an updated state. 
Given a statement $S$ and a context $c$,  we write $\flowftwo{S}{c}$ to denote
$\flowfthree{S}{c}{\sigma_S}$ by making $\sigma_S$ an implicit argument where
$$\sigma_S = join(c,S) $$

Recall from on our running example, the imprecision of
the context insensitive analysis is caused by the two calls of {\tt seq} in
lines 12 and 13 in Figure~\ref{f:get_cps_flatten}.
If we define the context to be last call sites of the function, i.e. program locations, we would
achieve a better precision,

{
  \begin{tabular} {  | c c c | c  c c | }
    \hline
  var & context & func & var & context & func \\ \hline
  {\tt first} & 12 & $ \lambda_{78}$ & {\tt second} & 12 & $\lambda_{35}$ \\
  {\tt first} & 13 & $ \lambda_{18}$ & {\tt second} & 13 & $\lambda_{68}$ \\
  \hline
\end{tabular}
}  
\\
This is also known as the context sensitive with call string. In the above case we use
a call string with size of 1. However in the presence of multiple loops in the source code,
the obfuscated code will contain multiple calls to the {\tt loop} combinator, which contains a recursion.
Choosing the size of the call string is a non-trivial task. A similar
observation applies to other context sensitive analyses, such as
functional approach, which consider the abstract state at the call site to be the context.
The worst case complexity of these context sensitive analyses makes them
less-practical to be applied in reverse engineering attacks without using heuristics \cite{spa}.

\subsection{Complexity of Sub-graph isomorphism}
Regardless of the precision of the static analysis result, 
it is computationally expensive to 
match the original control flow graph with the approximated control flow graph in general.
Let the original CFG to be $H$ and the approximated CFG to be $G$, we
want to check whether $H$ is sub graph isomorphic to $G$, which is
NP-complete \cite{Cook71thecomplexity}. For instance Ullmann's
algorithm \cite{Ullmann76analgorithm} is known to be exponential. Some
improvement with heuristic algorithms exist. There is no known
algorithms solving this problem in polynomial time.
This check only returns yes or no. Finding all possible isomorphic
sub-graphs leads to sub graph matching problem, which is also
NP-Complete.

Note that some linear algorithm exists for the special case in which one of the input
graphs is fixed and the other is a planar 
graph. Unfortunately CFG generated from Java in general is not
guaranteed to be a planar graph \cite{cfgplanarity}.

\ignore{
\section{Implementation} \label{sec:impl}
We report the implementation of the SSA to CPS translation in
Haskell \cite{cps-obs}. The
implementation provides a source-to-source obfuscation for C
programs. The prototype employs standard SSA construction found in
literature and converts the SSA into CPS. 
Since nested function definition, lambda expressions and closures are not supported in
C, we adopt the implementation tricks, such as lambda lifting,
function pointers and
continuation-as-a-stack, which were 
inspired by Kerneis and Chroboczek's work \cite{DBLP:journals/corr/abs-1011-4558}.

We benchmark our CPS obfuscation against CFF. Our preliminary
benchmark shows that CPS yields an average of 
3.96 times increase in source code size and 10.3 times in object code
size, compared to CFF, 1.52 times and 1.25 times respectively. 
Our benchmark also shows that the CPS obfuscated codes have an average
30\% run-time overhead and CFF obfuscated codes have an average 2.8\% overhead.
}

\section{Related Works} \label{sec:related-works}



The CPS-based control flow obfuscation is rooted from the connection
between SSA forms in imperative programming languages and lambda terms in
functional programming languages \cite{chakravarty03perspective,Appel:1998:SFP:278283.278285,Kelsey:1995:CCP:202529.202532}.
Our translation scheme is an extension of Lu's
work\cite{DBLP:conf/pepm/Lu19} and is
inspired by Kelsey's work
~\cite{Kelsey:1995:CCP:202529.202532}. In contrast with Lu's work, we
are targeting FJ instead of C style language.
As an improvement to Lu's work, our translation scheme supports
exception handling, recursive call and call to methods within the same scope with continuations.
In contrast to Kelsey's work, our translation is targeted at an imperative language extended
with higher order function instead of Scheme. 
Giacobazzi et al proposed a method to construct general obfuscators
using partial evaluation with distorted interpreters
\cite{Giacobazzi:2012:OPE:2103746.2103761}. Their work provides a
uniform reasoning of how attacks using abstract interpretation can be
foiled by a particular obfuscation method (by constructing a specific
distorted interpreter). 
Anonca and Corradi \cite{DBLP:conf/ecoop/AnconaC16} formalized SSA
form for FJ. They applied SSAFJ to improve the type
analysis of object oriented languages such as Java.

We note that it is possible that attackers are aware of this
obfuscation technique and try to reverse-engineer the obfuscation via
CPS, for instance, by applying the technique found in
\cite{DANVY1994183}. However this unclear to us that how much
additional information the attackers can recover by converting the
obfuscated code in CPS back to direct style. For instance, if we
ignore the treatment of exception, one could translate the {\tt loop}
combinator in CPS back to direct style as follows,

\begin{lstlisting}
void loop ( void => Boolean cond
          , void => void visitor,
          , void => void exit) { 
  if (cond()) {
    visitor();
    loop(cond, visitor, exit);
  } else {
    exit();
  }
}
\end{lstlisting}
Similar treatment can be applied to other obfuscated code in CPS
style. As observed, such translation does not improve the precision of
the static analysis.

\section{Conclusion} \label{sec:conclusion}

We extend and develop CPS-based control flow obfuscation for
FJ with exception handling.
We formalize the strategy as a source to source translation scheme.
We show that the control flow obfuscation technique is effective
against attacks using static control flow analysis, in particular
context insensitive analysis. We are in the process of implementing
the reported technique. The progress and some examples can be found in our development repository \cite{obsidian}.



\bibliography{main}

\appendix
\section{Appendix}

\subsection{Pre-processing step that fix while block that has multiple
  entry labels} \label{sec:fixing-preproc}

  The only possible case that violates the restrictive form is the use
  of try-catch with a while loop in the handler.
  \begin{lstlisting}
try {
  ...
  Li : throw new Exception();
  ...
  Lj : throw new Exception();
} join (...) catch (Exception e) {
  Ll:join (x = phi(Li:xi Lj:xj, Lm:xm)) while (e) {
    Lm: ...
  }
}
  \end{lstlisting}
  The above can be converted into the restrictive form by inserting an empty assignment block in front of the while
  block.
  \begin{lstlisting}
try {
  ...
  Li : throw new Exception();
  ...
  Lj : throw new Exception();
} join (...) catch (Exception e) {
  Lk: { };
  Ll: join (x = phi(Lk:xk, Lm:xm)) while (e) {
    Lm: ...
  }
}
  \end{lstlisting}

\end{document}